\documentclass[aps,prb,reprint,superscriptaddress,longbibliography]{revtex4-1}

\usepackage{xcolor}
\usepackage{textcomp}
\usepackage{graphicx} % for pdf, bitmapped graphics files
\usepackage{float}
\usepackage{epstopdf}
\usepackage{txfonts}

\usepackage[colorlinks,citecolor=blue,linkcolor=blue]{hyperref}

\newcommand{\UMPhy}{Unit\'e Mixte de Physique, CNRS, Thales, Universit\'e Paris-Saclay, 91767, Palaiseau, France} 
\newcommand{\SPEC}{SPEC, CEA, CNRS, Universit\'e Paris-Saclay, 91191, Gif-sur-Yvette, France}
\newcommand{\CNN}{Centre de Nanosciences et de Nanotechnologies, CNRS, Universit\'e Paris-Saclay, 91120, Palaiseau, France}

\begin{document}

\title{Resonant dynamics of skyrmion lattices in thin film multilayers: Localised modes and spin wave emission}

% [JVK] Prefer macros defined by RevTeX to define authors and affiliations
\author{Titiksha Srivastava}
\email{titiksha.srivastava@cea.fr}
\affiliation{\UMPhy}
\affiliation{\SPEC}
\author{Yanis Sassi}
\affiliation{\UMPhy}
\author{Fernando Ajejas}
\affiliation{\UMPhy}
\author{Aymeric Vecchiola}
\affiliation{\UMPhy}
\author{Igor Ngouagnia}
\affiliation{\SPEC}
\author{Herv\'e Hurdequint}
\affiliation{\SPEC}
\author{Karim Bouzehouane}
\affiliation{\UMPhy}
\author{Nicolas Reyren}
\affiliation{\UMPhy}
\author{Vincent Cros}
\affiliation{\UMPhy}
\author{Thibaut Devolder}
\affiliation{\CNN}
\author{Joo-Von Kim}
\email{joo-von.kim@c2n.upsaclay.fr}
\affiliation{\CNN}
\author{Gr\'egoire de Loubens}
\email{gregoire.deloubens@cea.fr}
\affiliation{\SPEC}

\begin{abstract}

The spectral signatures of magnetic skyrmions under microwave field excitation are of fundamental interest and can be an asset for high frequency applications. These topological solitons can be tailored in multilayered thin films, but the experimental observation of their spin wave dynamics remains elusive, in particular due to large damping. Here, we study Pt/FeCoB/AlO$_x$ multilayers hosting dense and robust skyrmion lattices at room temperature with Gilbert damping of $\sim 0.02$. We use magnetic force microscopy to characterise their static magnetic phases and broadband ferromagnetic resonance to probe their high frequency response. Micromagnetic simulations reproduce the experiments with accuracy and allow us to identify distinct resonant modes detected in the skyrmion lattice phase. Low ($<$ 2~GHz) and intermediate frequency ($2-8$~GHz) modes involve excitations localised to skyrmion edges in conjunction with precession of the uniform background magnetisation, while a high frequency ($>$ 12~GHz) mode corresponds to in-phase skyrmion core precession emitting spin waves into uniform background with wavelengths in the 50--80~nm range commensurate with the lattice structure. These findings could be instrumental in the investigation of room temperature wave scattering and the implementation of novel microwave processing schemes in reconfigurable arrays of solitons.

\end{abstract}

\maketitle

The dynamic response of magnetic materials at microwave frequencies represents a rich field of research for its fundamental interest and applications in information processing. Linear excitations in the form of spin waves (SWs) underpin the field of magnonics, which describes the paradigm of transmitting and processing information with such waves \cite{kruglyak10, krawczyk14, barman21}. SW-based devices may allow for fast and energy-efficient logic applications \cite{csaba17}, and a growing number of proposals have shown potential uses for computing \cite{macia11} and spectral analysis \cite{papp17} by SW interference. Magnonic circuit elements such as transistors \cite{chumak14}, diodes \cite{grassi20} and filters \cite{merbouche21} have also been demonstrated in experiments based on the manipulation of dipole-dominated SWs with micrometer wavelengths. The generation and detection of shorter wavelength SWs, a prerequisite for miniaturization, is challenging, owing to the limitations of nanoscale fabrication. Recent studies have shown that short wavelength emission can be achieved by broadband antennae \cite{ciubotaru16}, grating effects \cite{yu13, yu16, chen19} and spin torques \cite{fulara19}.

In this light, nonuniform magnetic textures have been explored for generating and manipulating spin waves, and generally offer an alternative route to expand the range of useful phenomena for applications \cite{yu21}. Such textures can appear spontaneously at the micro- and nano-scale in magnetic materials as a result of the competing interactions, namely, exchange, dipolar, and anisotropy. Magnetic textures like bubbles \cite{argyle83, montoya17}, stripes \cite{vukadinovic01, ebels01, gubbiotti12}, and vortices \cite{buess04, castel12} have been shown to exhibit a rich diversity of magnetisation dynamics. For example, it has been demonstrated that magnetic domain walls can serve as nanoscale waveguides \cite{garcia-sanchez15,wagner16}, while the cores of magnetic vortices can be used to generate omnidirectional dipole-exchange SWs with sub-100~nm wavelengths \cite{wintz16, dieterle19}.

Recently, topologically non-trivial chiral magnetic configurations called skyrmions have generated much interest owing to their robust and particle-like nature~\cite{fert17}. They can be stabilised at room temperature in thin films~\cite{moreau-luchaire16, boulle16} with perpendicular magnetic anisotropy (PMA) and interfacial Dzyaloshinskii-Moriya interaction (iDMI), which provides different handles for tuning the desired magnetic parameters, both statically and dynamically~\cite{srivastava18}. On one hand, their dc current-driven dynamics~\cite{jiang15, woo16} could be exploited for racetrack memory and logic devices \cite{fert17}, and on the other, their unique microwave response opens up the possibilities of skyrmion-based spin-torque oscillators~\cite{garcia-sanchez16}, rf detectors~\cite{finocchio15}, and reconfigurable magnonic crystals for microwave processing \cite{ma15, mruczkiewicz16, garst17, wang20}. Skyrmions exhibit a rich variety of eigenmodes~\cite{schuette14, kim14, zhang17, mruczkiewicz17, kravchuk18}, among which azimuthal bound states and breathing modes have been experimentally observed at low temperatures in bulk crystals~\cite{onose12, schwarze15, aqeel21}, which have low damping parameters. Only very recently, resonant dynamics with specific spectroscopic signatures of thin-film multilayers hosting skyrmions has been evidenced~\cite{satywali21, flacke21}. However, much remains to be explored and understood concerning their individual and collective resonant response to microwave excitations.

In this work, we report a study of the resonant dynamics of ultrathin film multilayers with perpendicular magnetic anisotropy, which host stable skyrmion lattices under ambient conditions with typical periods of 250~nm and skyrmion diameters of 100~nm, while exhibiting Gilbert damping in the range of $\alpha \simeq 0.02$. By combining magnetic force microscopy (MFM) and ferromagnetic resonance (FMR) experiments with micromagnetic simulations, we can identify distinct SW modes associated with the skyrmion lattice phase. At low frequency ($\lesssim$ 2~GHz), we observe a number of modes related to the precession of the uniform background state of individual layers close to or at the surfaces of the stack, along with eigenmodes localised to the skyrmion edges. At intermediate frequencies ($2-8$~GHz), the precession of the uniform background near the centre of the stack dominates the response. Similar modes were previously described and reported in bulk crystals~\cite{schwarze15, aqeel21}, and lately in thin films~\cite{satywali21, flacke21}. Intriguingly, we also observe a well-defined mode at high frequency ($>$ 12~GHz), which corresponds to the in-phase precession of the magnetisation within the skyrmion cores. The cores possess a distinct three-dimensional structure due to the competition between all the existing magnetic interactions in these multilayers, notably the interlayer dipolar effects~\cite{legrand18}. Strikingly, this precession is accompanied by the emission of spin waves, with wavelengths in the range of 50 to 80 nm, into the uniformly magnetised background. These SWs interfere with those generated at neighbouring skyrmion cores, yielding a collective dynamical state governed by the subtle interplay between the skyrmion diameter, the wavelength of the emitted SWs, and the skyrmion lattice periodicity.

\section*{Results}

\subsection*{Multilayer composition}

The basic element of the multilayer film studied is the Pt(1.6)/Fe$_{0.7}$Co$_{0.1}$B$_{0.2}$(1.2)/AlO$_x$(1.0) trilayer (hereafter referred to as Pt/FeCoB/AlO$_x$), where the figures in parentheses indicate the nominal film thickness in nm. This trilayer lacks inversion symmetry along the film thickness direction whereby the Dzyaloshinskii-Moriya interaction is promoted at the interfaces of the ferromagnetic FeCoB film with Pt (and possibly AlO$_x$ \cite{yang18}). The trilayer is repeated 20 times to form our multilayer sample (see Methods), as shown in Fig.~\ref{fig:loopMFM}(a). The choice of Fe-rich FeCoB and the aforementioned optimised thicknesses of Pt and FeCoB allow having sufficient DMI and PMA \cite{kim17a, tacchi17, legrand21} to stabilise skyrmions, while limiting spin pumping effects which would otherwise lead to an increased damping coefficient \cite{belmeguenai18}. We estimate by FMR a Gilbert damping constant of $\alpha = 0.022$ and an inhomogeneous broadening of $\sim$ 18~mT (see \textcolor{blue}{Supplementary Figure 1}) for our samples, which are relatively low for such multilayer systems~\cite{satywali21, belmeguenai18}. The overall magnetic volume is enhanced with the 20 repetitions of the trilayer, which not only increases the thermal stability of the skyrmions \cite{moreau-luchaire16} but also provides a larger signal-to-noise ratio for inductive measurements.

\subsection*{Static characterization}

\begin{figure}
	\centering
        \includegraphics[width=\columnwidth]{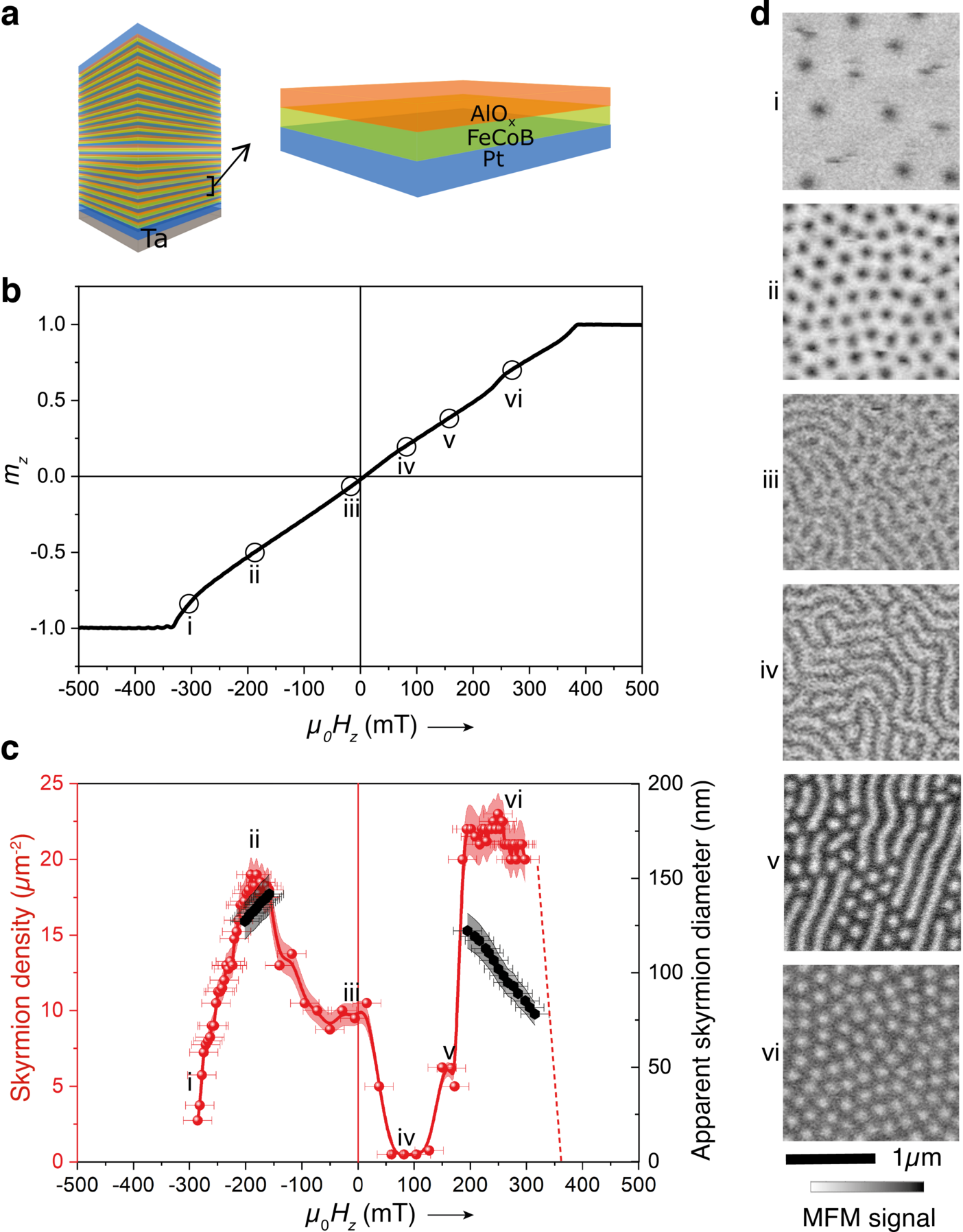}
	\caption{\textbf{Static characterization of the magnetic phases.} (a) Schematics of the sample stack. (b) The out-of-plane (OP) magnetisation curve on sweeping the field from negative to positive values. (c) The corresponding skyrmion density and apparent skyrmion diameter as a function of the field. The dashed red line indicates the possible extrapolation of the observed skyrmion density at higher positive field values in accordance with the OP hysteresis curve. The skyrmion diameter is extracted only for the field ranges where the skyrmion density is close to maximum (see Methods). (d) Magnetic domain configurations of the sample: (i) randomly distributed skyrmions, (ii and vi) skyrmion lattice, (iii and v) skyrmion-stripe mix, and (iv) labyrinthine domains, measured by MFM as a function of OP magnetic field (corresponding to the values marked in (b)) swept from negative to positive values. }
	\label{fig:loopMFM}
\end{figure}

Figure~\ref{fig:loopMFM}(b) shows the out-of-plane hysteresis curve of the sample measured by alternating gradient field magnetometer (AGFM), which is characteristic of thin films hosting skyrmions. The quasi-static magnetisation configuration of the sample is probed by MFM using a low moment tip by sweeping the out-of-plane (OP) field from negative ($-z$) to positive ($+z$) saturation as illustrated in Fig.~\ref{fig:loopMFM}(d). On decreasing the field from negative saturation, we observe nucleation of skyrmions with an average diameter around 100~nm (zone (i)), which develops into a dense lattice structure upon further reduction of the field (zone (ii)). In zone (iii) some of the skyrmions elongate and/or coalesce to form stripes leading to a mixture of skyrmions and stripe domains. At small positive fields, labyrinthine domains (zone (iv)) become energetically stable and, upon increasing the field further, lead to the formation of a dense skyrmion lattice once again (zone (vi)). From the MFM images we extract the density of skyrmions as a function of the applied field $\mu_0H_z$ and apparent skyrmion diameter for the ranges of magnetic field where the skyrmion density is close to maximum, as shown in Fig.~\ref{fig:loopMFM}(c).

It is interesting to note that the observed skyrmion lattice is almost quasi periodic hexagonal (see \textcolor{blue}{Supplementary Figure 2}) whereas more amorphous states were observed in previous studies~\cite{satywali21, flacke21} in such kind of multilayered films, indicating only minor inhomogeneities and defects induced during the film growth in our samples. The skyrmion lattice phase is also remarkably stable at positive field, as shown by the constant skyrmion density of about 21 per square micron between 200~mT and 350~mT.

By using the period of the labyrinthine domain pattern in Fig.~\ref{fig:loopMFM}(d)(iv) and the measured values of saturation magnetisation ($M_s=1.2$~MA/m) and uniaxial anisotropy ($K_u=0.7$~MJ/m$^3$), we estimate the DMI to be $D=1.2$~mJ/m$^2$ with an exchange constant of $A=15$~pJ/m for our sample (see Methods), which are in good agreement with direct experimental determination of these parameters in Pt/FeCoB/MgO stacks \cite{boettcher21}.

\subsection*{Dynamic characterization}

\begin{figure*}
	\centering
        \includegraphics[width=12cm]{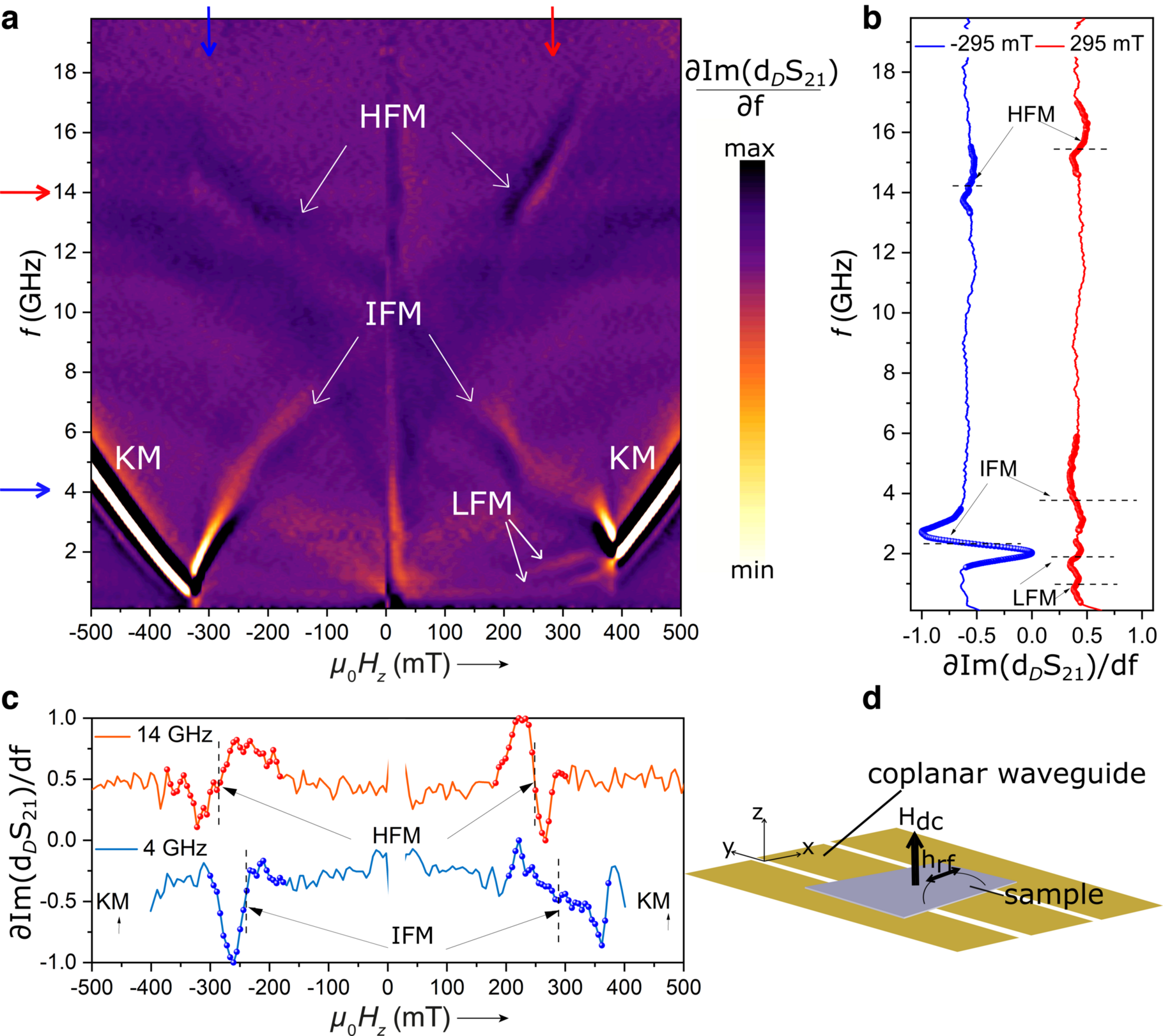}
	\caption{\textbf{Broadband ferromagnetic resonance.} (a) Frequency-field dispersion map measured by VNA-FMR over a frequency range of 0.1~GHz to 20~GHz with the applied OP field swept from -500~mT to 500~mT. Apart from the Kittel mode (KM) above saturation, several distinct modes appear in the non-saturated state which are labelled as HFM: high frequency mode ($f> 12$~GHz), IFM: intermediate frequency mode ($f< 8$~GHz) and LFM: low frequency modes ($f< 2$~GHz). (b) Line cuts for fixed OP field values: -295~mT and 295~mT. (c) Line cuts for fixed frequency values: 4~GHz and 14~GHz. (d) Schematic of the VNA-FMR setup.}
	\label{fig:FMRmodes}
\end{figure*}

The resonant dynamics of the sample was probed by broadband FMR using a coplanar waveguide [Fig.~\ref{fig:FMRmodes}(d)] and a vector network analyser (VNA). The dc out-of-plane field is swept from negative to positive values while the frequency of the in-plane ac field is scanned over a range of 0.1--20~GHz for each dc field step. The real and the imaginary parts of the transmission signal $S_{21}$ are recorded and processed to remove a background signal that is independent of the dc field, which improves the contrast (see Methods and \textcolor{blue}{Supplementary Figure 3}). The corresponding frequency versus field map is shown in Fig.~\ref{fig:FMRmodes}(a). In the saturated state, we observe the high intensity Kittel mode (KM) along with two additional low intensity secondary modes at higher field (see also \textcolor{blue}{Supplementary Figure 1}). The latter are attributed to localised modes in the multilayer thickness that result from inhomogeneous interfacial couplings of our multilayer system \cite{puszkarski94}. In addition to the KM observed above saturation fields, three groups of modes with lower intensities appear when the magnetisation enters a non-uniform state. On ramping the field from negative saturation towards zero, the KM softens close to $\mu_0H_z=-335$~mT where skyrmions start to nucleate (see MFM images in Fig.~\ref{fig:loopMFM}(d)). It then evolves into a mode with negative field dispersion (\textit{i.e.}, $\partial f/\partial|H_z|<0$) as the skyrmion lattice grows denser, which we call the intermediate frequency mode (IFM). At this point, a weak amplitude mode also emerges at high frequency ($> 12$~GHz) which has a positive field dispersion (\textit{i.e.}, $\partial f/\partial|H_z|>0$) denoted as the high frequency mode (HFM). The IFM and HFM fade away close to zero field, where the static magnetisation configuration consists of labyrinthine domains. While the IFM reappears as the field is increased towards positive values corresponding to a mixture of skyrmions and stripes, the HFM is visible only above 200~mT when the magnetisation profile consists of a dense skyrmion lattice network (see Fig.~\ref{fig:loopMFM}(c)), and the IFM dispersion nearly flattens. We also notice that at the very same field a mode appears at low frequency ($<2$~GHz) followed by another at even lower frequencies, which we refer to as low frequency modes (LFM). The HFM mode fades away at fields when the skyrmion lattice transforms into isolated skyrmions. The IFM continues beyond this point however with an abrupt change of slope until the magnetisation becomes uniform beyond 390~mT. The line cuts along fixed fields ($-295$~mT and $295$~mT) and fixed frequencies (14~GHz and 4~GHz) are shown in Figs.~\ref{fig:FMRmodes}(b) and \ref{fig:FMRmodes}(c) respectively. The modes at negative and positive fields are not symmetric but instead depend on the magnetic field history (see \textcolor{blue}{Supplementary Figure 4}), as does the static magnetisation profile.

\subsection*{Micromagnetic simulations}

\begin{figure*} 
	\centering
        \includegraphics[width=16cm]{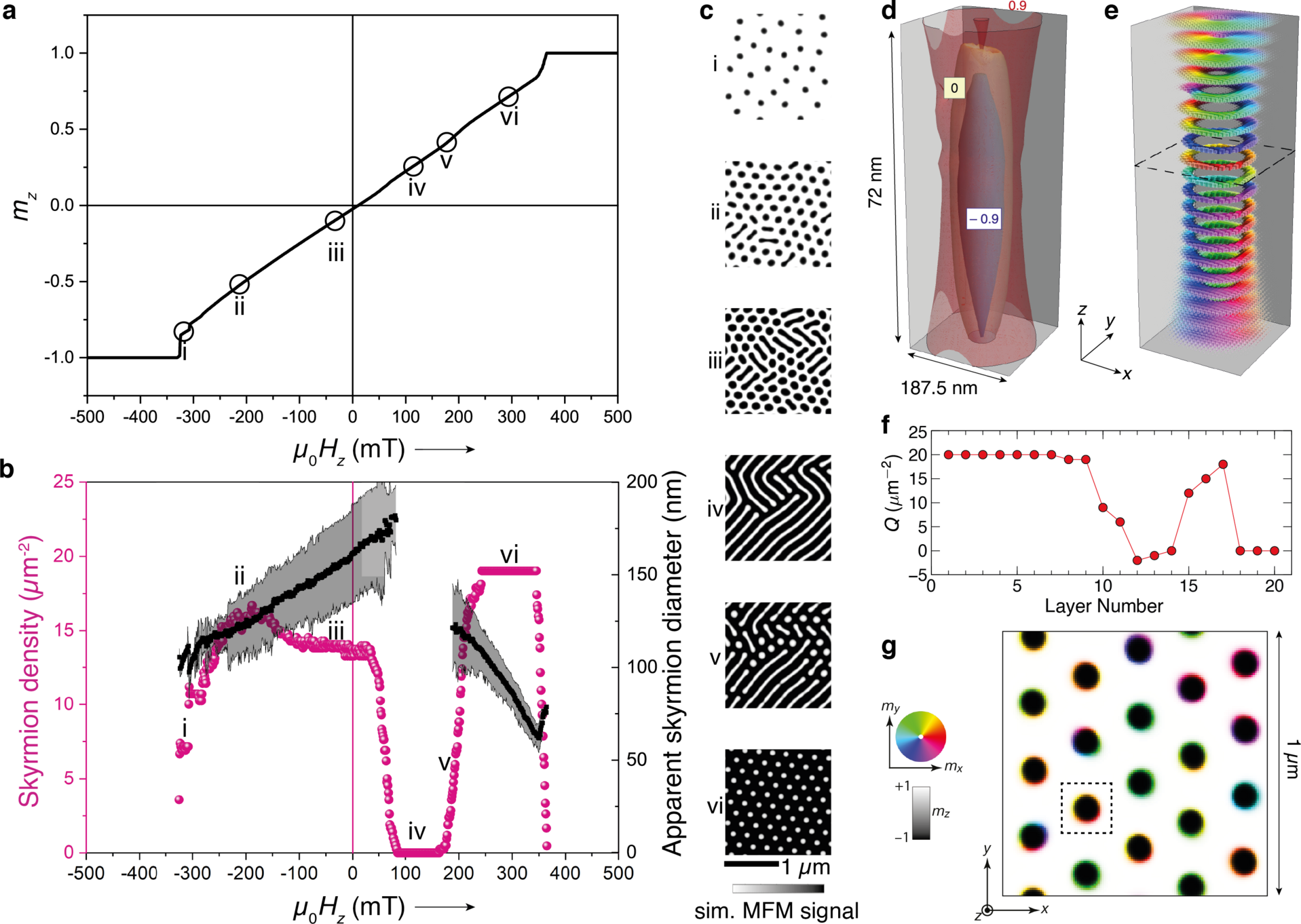}
	\caption{\textbf{Micromagnetic simulations of equilibrium magnetic configuration.} (a) Simulated magnetisation curve, (b) skyrmion density and diameter evolution and (c) the corresponding simulated MFM images for the OP field swept from negative to positive values. The grey shaded area in (b) results from uncertainties in determining the skyrmion diameter, as discussed in the Methods section. (d) $m_z$ contours and (e) in-plane components of the magnetisation texture of a single skyrmion in the lattice phase at 285~mT. (f) Topological charge density $Q$ per layer 285~mT. (g) Magnetic configuration of layer 12 at 285~mT.}
	\label{fig:Simstat}
\end{figure*}

We performed simulations with the finite-difference micromagnetic code \textsc{Mumax3} \cite{vansteenkiste14} in order to gain better insight into the static and dynamic properties of our sample (see Methods). We modelled the full 20-layer repetition of the Pt/FeCoB/AlO$_x$ trilayer (with periodic boundary conditions in the film plane) in order to account for dipolar effects as accurately as possible, since it is known that the skyrmion core deviates from the usual tubular structure to complex three-dimensional configurations due to inhomogeneous dipolar fields along the multilayer thickness~\cite{legrand18}. Figures~\ref{fig:Simstat}(a) and \ref{fig:Simstat}(b) show the simulated hysteresis loop and the skyrmion density and apparent diameter variation as a function of the OP field swept from negative to positive values. The MFM images calculated from the simulated magnetisation profiles (see Methods) are shown in Fig.~\ref{fig:Simstat}(c). The field evolution of the simulated static characteristics presented in Figs.~\ref{fig:Simstat}(a)--(c) is found to be in good agreement with the experiments.

A striking feature of the labyrinthine domain and skyrmion structures found is that their micromagnetic configuration exhibits strong variations along the multilayer thickness direction. An example of such complex structures is shown in Fig.~\ref{fig:Simstat}(d), where the $m_z$ component is shown for a single skyrmion core in the lattice phase at $\mu_0 H_z =285$~mT. The figure shows the contours for $m_z = 0.9$ (red), $m_z = 0$ (yellow), and $m_z = -0.9$ (blue) as surfaces where cubic interpolation has been used across the nonmagnetic layers. We note that the skyrmion core, in particular the region of reversed magnetisation $m_z \leq -0.9$, does not extend across the entire thickness of the multilayer. Recall that the FeCoB layers are only coupled together through dipolar interactions, which are sufficiently large to maintain an alignment of the core centre but too weak to promote a coherent magnetisation profile across the different layers. We can also observe that the magnetisation in the uppermost layers at the core centre is not reversed at all, but slightly tilted away from the film normal as indicated by the presence of the inverted cone. The in-plane components of the magnetic texture shown in Fig.~\ref{fig:Simstat}(d) are presented in Fig.~\ref{fig:Simstat}(e). Here, each cube represents the magnetic state of a finite-difference cell, where the colour represents the orientation of the magnetisation in the cell. In order to highlight the role of the in-plane components to complement the data shown in Fig.~\ref{fig:Simstat}(d), the relative size and opacity of each cube is scaled with the function $1-m_z^2$; this renders the regions of the uniform background and the reversed magnetisation transparent. A clear skyrmion profile can be seen for the nine bottom layers, where the same left-handed (\textit{i.e.} counterclockwise) N{\'e}el chirality is found in each layer. In layers 10 to 14, on the other hand, the in-plane component of the magnetisation at the skyrmion boundary becomes more uniform and does not exhibit the same winding as in the bottom half of the stack. This means that the reversed domain structure is non-topological.  A skyrmion profile reappears in layers 15, 16 and 17 but with a reversed chirality, where the in-plane magnetisation components of the right-handed (\textit{i.e.} clockwise) N{\'e}el structure are rotated by 180 degrees in the plane with respect to their left-handed counterparts. Finally in layers 18 to 20, we observe another type of non-topological texture where the core magnetisation is closely aligned with the background magnetisation, which corresponds to the inverted cone at the top of the stack in Fig.~\ref{fig:Simstat}(d). Figure~\ref{fig:Simstat}(f) shows the variation of the topological charge density per layer as a function of the layer number, which shows that a similar thickness dependence is observed across the skyrmion lattice. The skyrmions in the bottom half of the stack remain topological, while the top half comprises largely non-topological bubbles. Finally, Fig.~\ref{fig:Simstat}(g) illustrates the magnetic configuration of layer 12 across the entire region simulated, where we can observe that the mainly uniformly-magnetised regions of the magnetic bubble walls can vary greatly from one bubble to the next, with no discernible spatial order. We have verified that these features persist for finite difference cell sizes down to $\sim$ 2 nm, which indicates that the complex magnetisation structure does not arise from discretization effects (see \textcolor{blue}{Supplementary Figure 5}).

\begin{figure*} 
	\centering
        \includegraphics[width=16cm]{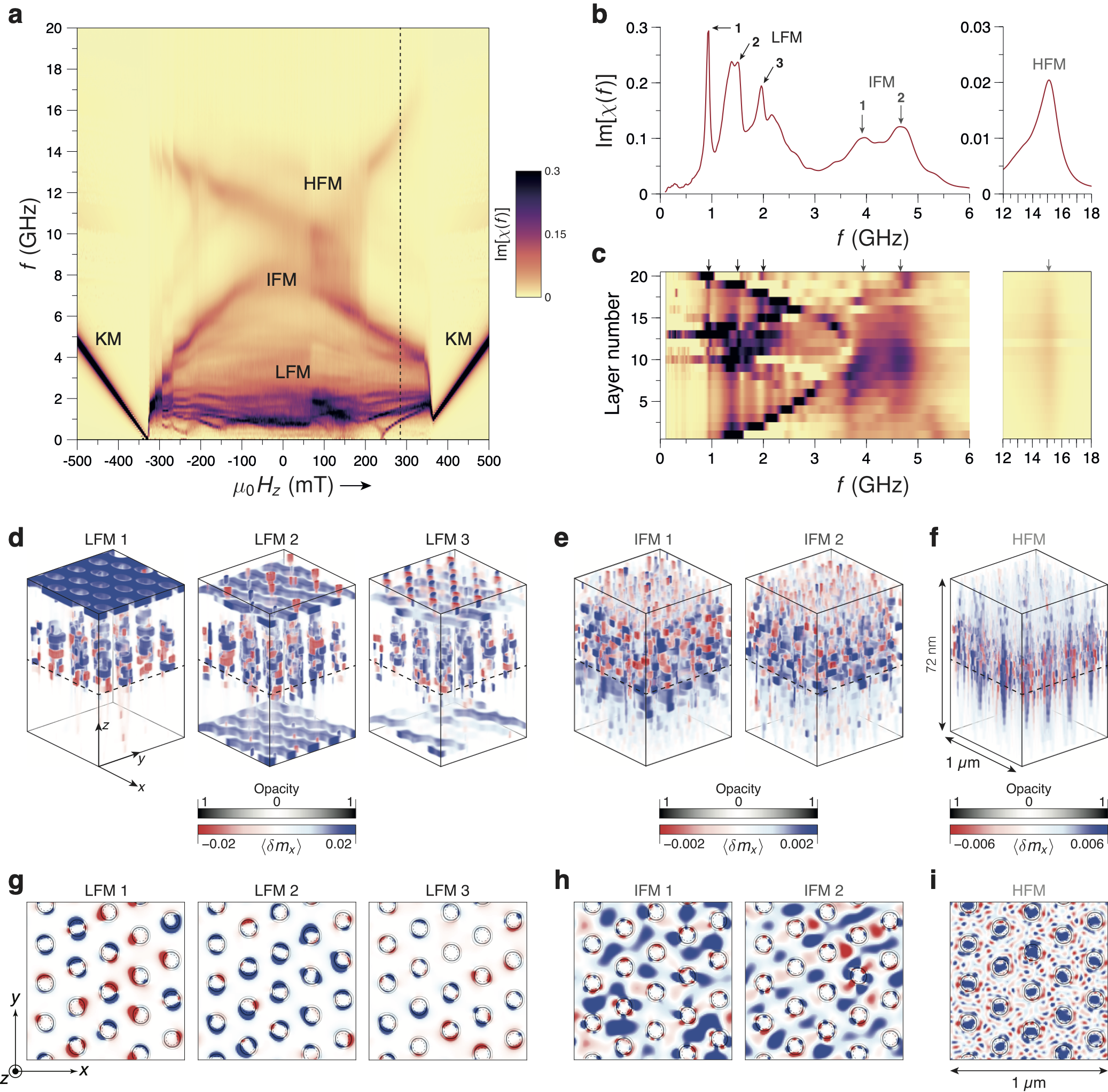}
	\caption{\textbf{Micromagnetic simulations of the dynamic response.} (a) FMR spectra simulated by calculating the frequency-dependent susceptibility of multilayer to uniform excitation fields. (b) Imaginary part of the magnetic susceptibility at 285~mT (indicated by a dashed line in (a)), showing three LFM (0.94, 1.50, and 1.92 GHz), two IFM (3.90 and 4.24 GHz) and one HFM (15.10 GHz). (c) Thickness-resolved susceptibility at  285~mT. (d, e, f) Three-dimensional view of the magnetic response $\langle\delta m_x\rangle$ for (d) LFM, (e) IFM, and (f) HFM. (g, h, i) Two-dimensional view of the magnetic response $\langle\delta m_x\rangle$ in layer 10 of the modes in (d,e,f) for (g) LFM, (h) IFM, and (i) IFM.}
	\label{fig:Simdyn}
\end{figure*}

We next discuss the dynamical response of the system, where the frequency-dependent susceptibility is computed under different applied fields as in Fig.~\ref{fig:FMRmodes}(a) (see Methods). The simulated susceptibility map is shown in Fig.~\ref{fig:Simdyn}(a), which is determined from response of the static configurations computed in Fig.~\ref{fig:Simstat} to sinusoidal in-plane fields in the frequency range of 0.1 to 20~GHz. The Kittel mode (KM) is easily identified for the uniform state for fields above the saturation field. In the regime in which the magnetisation is nonuniform,$-330 < \mu_0H_z< 365$~mT, three distinct types of modes can be identified, as illustrated by the line cut at 285~mT shown in Fig.~\ref{fig:Simdyn}(b). As in experiments, the KM transforms into a negative field dispersion mode, an intermediate frequency mode (IFM), at fields where skyrmion nucleation begins. The IFM dispersion is rather rugged on the negative field side, where the skyrmion density rapidly evolves with field, and a faint splitting of IFM is seen around $-260$~mT. The IFM is asymmetric with respect to zero field and exhibits smoother variations on the positive field side. Similar to the negative field side, another branch of the IFM is seen for positive fields appearing at around $200$~mT, which is relatively flat around $4$~GHz in the range of positive magnetic fields $240 \leq \mu_0H_z\leq 345$~mT, where a dense skyrmion lattice is stable. It then merges to a single IFM and then varies sharply until saturation. Several low frequency modes (LFMs), comprising several closely spaced branches in the frequency range of $0.5<f<2.5$~GHz, extend over the entire magnetic field range in which the magnetisation is non-uniform. Finally, a high frequency mode (HFM) is also observed above 10~GHz with a positive field dispersion. It exhibits a sawtooth-like variation at negative fields and a smoother variation for positive fields. It is attenuated in the field range where magnetisation profile consists of randomly oriented stripes (region (iv) in Fig.~\ref{fig:Simstat}). The simulated susceptibility map reproduces well the three families of modes observed experimentally in the non-saturated state and gives good quantitative agreement for the mode frequencies. Note that in experiments, the LFM branches could not be resolved for the negative field side in the $S_{21}$ transmission data of Fig.~\ref{fig:FMRmodes}(a) (as opposed to the $S_{11}$ and $S_{22}$ reflection data, presented in \textcolor{blue}{Supplementary Figure 6}). It is because these modes have a nearly flat frequency--field dispersion, and hence get diluted in the signal processing which is necessary to remove the field independent offset. Only two of the LFM branches become resolvable when they acquire a certain slope on the positive field side, where the skyrmion density is constant and the skyrmion size strongly varies with the field.

The spatial profiles of the excited modes are found by driving the system at the mode frequencies identified in Fig.~\ref{fig:Simdyn}(a) and by recording the resonant response. In Fig.~\ref{fig:Simdyn}(b), we present the frequency response for $\mu_0 H_z = 285$ mT at which the equilibrium ground state is a skyrmion lattice, where we can identify well-defined peaks corresponding to the LFM, IFM, and HFM. Fig.~\ref{fig:Simdyn}(c) gives layer-resolved susceptibility whose sum over all layers results in the curve in Fig.~\ref{fig:Simdyn}(b). For the LFM, we observe that the peaks in Fig.~\ref{fig:Simdyn}(b) correspond to two kinds of resonances. For LFM 1 at $f = 0.94$ GHz, the primary response arises from the central part of the stack in layers 13 and 14, and also from the top of the stack in layer 20. A similar behaviour is seen for LFM 2 at 1.50 GHz, where layers 1 and 19, close to the stack surfaces, resonate along with the central part of the multilayer, as with LFM 3 at 1.92 GHz, where a strong response is seen in layers 2 and 18. This trend continues as the frequency increases, where successive layers farther away from the surfaces exhibit local resonances in conjunction with the central part. These mode profiles are visualised in Fig.~\ref{fig:Simdyn}(d), where we show the resonant fluctuations in the $x$ component of the magnetisation, $\langle \delta m_x \rangle$ (see Methods). Blue regions give the main response to the susceptibility, while the sum of equally intense blue and red regions cancel each other out. Similarly to Fig.~\ref{fig:Simstat}(e), the opacity of the cells increase with the strength of the response in order to provide a clearer picture of their spatial profiles. LFM 1 mainly corresponds to the uniform precession of the background magnetisation at the top of the multilayer stack, while the skyrmion edges are also excited in the central part of the stack with a small overall contribution to the susceptibility. This can be seen more clearly in Fig.~\ref{fig:Simstat}(g), where the fluctuations are shown for layer 10. For LFM 2 and LFM 3, a similar behaviour is seen where individual layers precess uniformly close to the surfaces, while excitations in the central part are localised to skyrmion edges.

For the IFM, two broad resonances are identified at $\mu_0 H_z = 285$ mT and correspond to excitations that are mainly localised to the central part of the multilayer [Fig.~\ref{fig:Simdyn}(c,e)]. The main contribution comes from the uniform precession of the background magnetisation between the skyrmion cores, i.e., the inter-skyrmion region which is aligned parallel to the applied field, which can been seen from the predominantly blue regions in Fig.~\ref{fig:Simdyn}(h) that appear outside the cores. 
%Note that the IFM branches relatively flatten at positive fields where the skyrmion density is constant, reflecting their pronounced dependence on the spin texture and hence the magnetic field history. 
In addition, some SW channelling is also observed at the skyrmion edges, as shown by alternating red and blue patches encircling the skyrmions.

The HFM involves the in-phase precession of the reversed magnetisation within the skyrmion cores and extends across most of the multilayer [Fig.~\ref{fig:Simdyn}(c,f)], with much of the power concentrated in the central part of the stack. The coherence of the precession can be seen in Fig.~\ref{fig:Simdyn}(i) by the uniform blue zones of $\langle \delta m_x \rangle$ within the $m_z = 0$ contours denoted by the continuous black circles. Moreover, this precession results in the generation of short wavelength SWs in the inter-skyrmion region, which are almost as intense as the core precession. However, only the volume occupied by the skyrmion cores provides a coherent response to the susceptibility for the HFM, which explains its much weaker intensity compared to the IFM and KM.

These features are also present at other applied field values at which the equilibrium state comprises nonuniform magnetisation. \textcolor{blue}{Supplementary Figure 7} shows an analogous behaviour at $\mu_0 H_z = -285$ mT, where the lattice possesses a lower skyrmion density with deformed core profiles. At $\mu_0 H_z = 140$ mT where the equilibrium state comprises labyrinthine domains, only the LFM and IFM are present, which further highlights the link between the HFM and the precessing skyrmion cores (\textcolor{blue}{Supplementary Figure 8}). At zero field, the equilibrium state comprises a mixture of bubbles and elongated stripes in which no clear background orientation can be defined. Here, LFM persist while the IFM and HFM are absent. Instead, we can observe the existence of cavity modes that represent confined spin waves (\textcolor{blue}{Supplementary Figure 9}).

\subsection*{Spin wave emission from the high frequency mode}

\begin{figure*} 
	\centering
        \includegraphics[width=14cm]{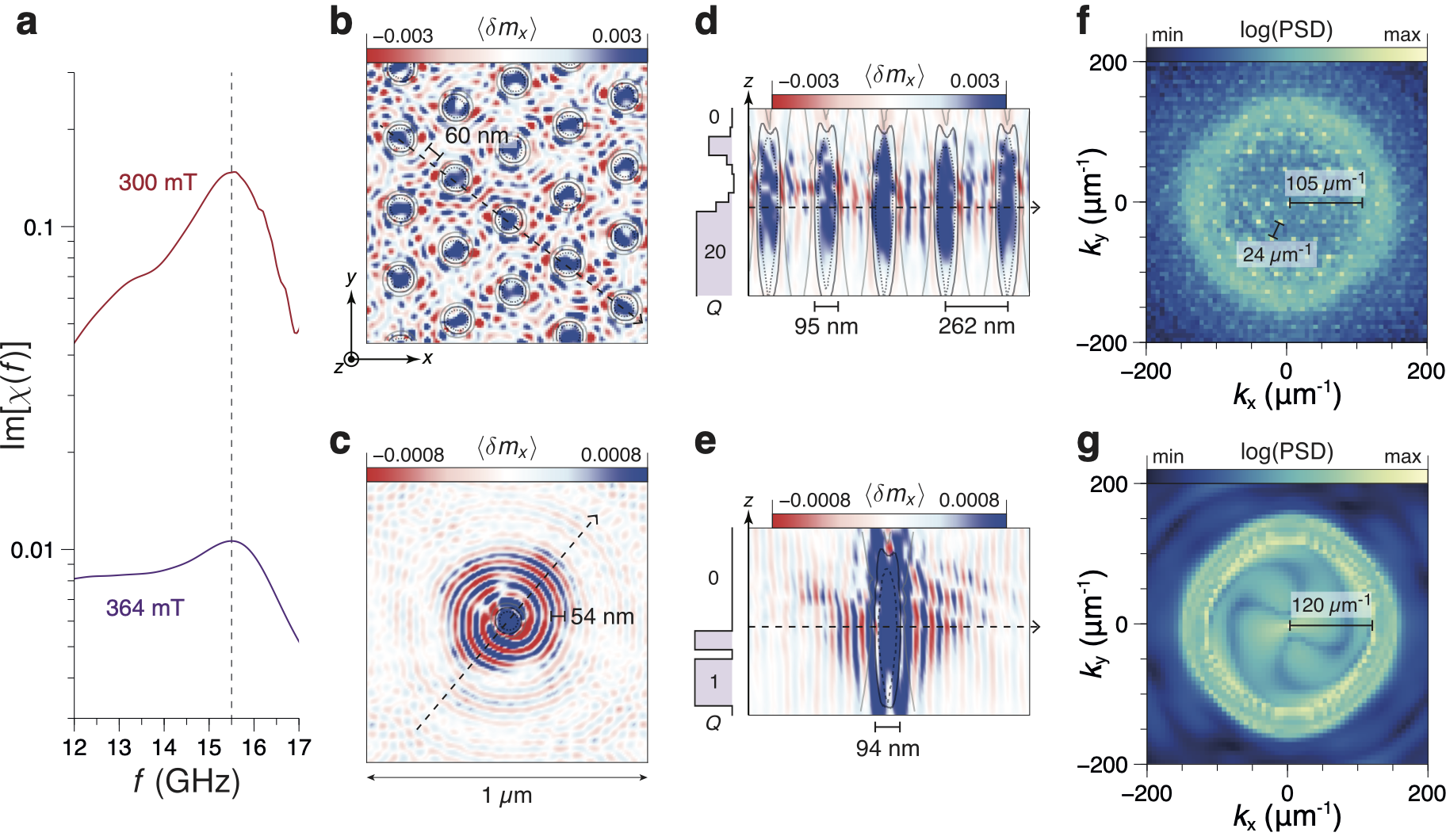}
	\caption{\textbf{Emission and interference of spin waves in skyrmion lattice.} (a) Simulated susceptibility curves at 300~mT and 364~mT with the dynamic magnetisation profiles $\langle\delta m_x\rangle$ in (b) and (c) of layer 10, respectively, corresponding to the resonance at 15.5~GHz. Cross-sections of $\langle\delta m_x\rangle$ marked by dotted lines in (b) and (c) are also shown respectively in panels (d) and (e), highlighting the in-phase skyrmions core precession (in blue) and spin wave emission in the surrounding medium (alternate blue and red zones) (f,g) Power spectral density (PSD) of the layer averaged spatial discrete Fourier transform of the dynamic magnetisation profiles indicating the average k$_{SW}$-vector of the emitted SWs at 300~mT and 364~mT respectively. At 300~mT, the skyrmion lattice parameter can also be deduced from the $k = 24~\mu$m$^{-1}$ periodic pattern seen in the spatial Fourier transform.}
	\label{fig:SimHFM}
\end{figure*}

The short wavelength SWs associated with the HFM cannot be excited directly with the uniform driving fields used in FMR, so their presence indicates the skyrmion cores act as nanoscale transducers that convert a spatially-uniform drive into a spatially-nonuniform excitation. To obtain deeper insight into this process, we focus on the HFM in the simulation data at two values of the applied fields, 300 and 364~mT, as presented in Fig.~\ref{fig:SimHFM}. The equilibrium state at 300~mT comprises a skyrmion lattice, with a skyrmion density of 20 $\mu$m$^{-2}$ and an average core-to-core distance of $L=262$~nm, while only one metastable skyrmion is present at 364~mT within the same area. The skyrmion diameters are roughly equivalent at these field values, which are 95 and 94~nm, respectively. The susceptibility at these two fields, shown in Fig.~\ref{fig:SimHFM}(a) for the range of 12--17~GHz, display a clear resonance peak around 15.5~GHz, which is over an order of magnitude larger at 300~mT than at 364~mT due to the difference in the skyrmion density.  %The resonant response observed at 250~mT has already been described in Fig.~\ref{fig:Simdyn}, with LFMs at 0.8 and 1.2~GHz, the IFM at 3.4~GHz, and the HFM at 12~GHz. At 330~mT, only two main resonances can be observed. The most intense peak occurs at 0.4~GHz, which corresponds to the uniform precession of the background where the magnetisation is oriented parallel to the OP field. Interestingly, a clear resonance peak also occurs at 12~GHz, which is at the same frequency as the HFM at 250~mT.
%Detailed inspection of the mode profiles extracted at this frequency and both fields, as shown in  Figs.~\ref{fig:SimHFM}(b)--(e), reveal that they correspond to the same HFM eigenmode. 

At both field values, the coherent precession of the reversed magnetisation within the skyrmion cores emit spin waves into the ferromagnetic background. Their wavelengths are similar at both fields and are estimated from the $\langle\delta m_x\rangle$ profiles in layer 10 shown in Figs.~\ref{fig:SimHFM}(b) and \ref{fig:SimHFM}(c), yielding $\lambda \simeq 60$~nm and $\lambda \simeq 54$~nm at $\mu_0H_z=300$~mT and $\mu_0H_z=364$~mT, respectively. Despite the strong variations in the static core profile across the multilayer thickness, the emitted spin waves are remarkably coherent across the thickness. This can be seen in Fig.~\ref{fig:SimHFM}(d) for 300~mT and Fig.~\ref{fig:SimHFM}(e) for 364 mT where coherent wavefronts, driven in the central part of the multilayer at which the core diameter is greatest, extend across much of the multilayer thickness. The left inset in Figs.~\ref{fig:SimHFM}(d,e) indicate the number of topological charges per layer, which shows that the spin wave emission is just as effective for the topological bubbles residing primarily in the bottom of the stack as for the non-topological bubbles residing primarily in the top. 

The overall coherence of these spin wave excitations can also be seen from the layer-averaged spatial discrete Fourier transform of the dynamical magnetisation shown in Figs.~\ref{fig:SimHFM}(b,c). These Fourier maps, shown in Figs.~\ref{fig:SimHFM}(f,g), display a ring pattern with wave vectors $100 < k_\mathrm{SW} < 150~\mu$m$^{-1}$ corresponding to wavelengths $40 < \lambda < 60$~nm. At 300~mT, the pattern in the Fourier transformed data also manifests the periodic structure of the skyrmion lattice, from which the lattice periodicity can be estimated as shown in Fig.~\ref{fig:SimHFM}(f), using $2\pi/k =262$~nm, where $k = 24~\mu$m$^{-1}$.

The spin wave dynamics can be understood from the linearised equations of motion for the dynamic magnetisation $\delta\mathbf{m}$, $\partial_t \delta\mathbf{m}=-|\gamma_0| (\delta\mathbf{m} \times \mathbf{H}_{\mathrm{eff},0} + \mathbf{m}_0 \times \delta \mathbf{H}_\mathrm{eff})$, where $\mathbf{m}_0$ is the static magnetisation configuration and $\mathbf{H}_{\mathrm{eff},0}$ is the static effective (internal) field associated with $\mathbf{m}_0$, while $\delta \mathbf{H}_\mathrm{eff}$ is the dynamic effective (internal) field associated with $\delta\mathbf{m}$. The full 3D structure of the core and the surrounding configuration determines the spatial profile of the static internal field and the localisation of the mode. The average value of this static internal field $\mathbf{H}_{\mathrm{eff},0}(\mathbf{r})$ inside the skyrmion core can be extracted from micromagnetic simulations and is about $-200$~mT, which leads to resonances at 5.8~GHz without any confinement effects. To this, one needs to add the dynamic dipolar and exchange contributions $\delta \mathbf{H}_\mathrm{eff}(\mathbf{r},t)$ due to the localisation of magnetisation dynamics in a volume of $\sim 100 \times 100 \times50$~nm$^3$, which increases the resonant frequency to about 15~GHz (see \textcolor{blue}{Supplementary Note 2}). 
%The precessing magnetisation in the skyrmion core generates a highly localised microwave field in its surroundings, which can efficiently couple to short wavelength SWs. 
Similarly, we can estimate the wavelength of the emitted SWs by using the dispersion relation of forward-volume SWs in the inter-skyrmion region. At 300~mT, the average internal field in this region is around 110~mT, which results in a wavelength of 60~nm at 15.5~GHz, a group velocity of 960~m/s and an attenuation length of 490~nm (see \textcolor{blue}{Supplementary Figure 10}). These estimations are in good agreement with the SW characteristics observed in Fig.~\ref{fig:SimHFM}. %Figs.~\ref{fig:SimHFM}(b) and \ref{fig:SimHFM}(f).

The difference between the mode profiles shown in Figs.~\ref{fig:SimHFM}(b) and \ref{fig:SimHFM}(c) is due to interference effects. At low skyrmion densities, the SW emitted at one core attenuates before reaching another, resulting in the familiar concentric ripples around each excitation source. At higher skyrmion densities, on the other hand, the inter-skyrmion distance is smaller than the attenuation length and interference patterns can arise. %Interestingly, this collective interference effect has repercussions on the amplitude of susceptibility at resonance. The spectral weight at 250~mT is 19 times larger that at 330~mT, whereas the skyrmion density is only 8.5 times larger. Hence the added contribution to the signal comes from the constructive interference of SWs in the skyrmion lattice. This matching is however tricky, it depends on several factors that have to be in line for this net constructive SW interference. 
The cores can emit SWs in a broad spectral range, which depends on the excitation frequency and the skyrmion size (see \textcolor{blue}{Supplementary Figure 11}). As such, the overall collective behavior is governed by a subtle interplay between the characteristic length scales of the system. The skyrmion diameter, which varies with the applied magnetic field, determines the precession frequency of the core thereby the wavelength of the emitted SWs, whereas constructive or destructive SW interference patterns are determined by the commensurability of SW wavelength and lattice parameters.

\section*{Discussion}
SW emission related to the HFM can have advantages over other reported methods of sub-100~nm wavelength generation. For example, magnetic vortex cores, which can act as SW emitters, require confined geometries such as dots for their existence and do not spontaneously form lattice structures in continuous films, which precludes the appearance of SW interference phenomena seen here. The precession frequency of the skyrmions in our samples is also above the FMR gap of the surrounding domain background and hence SWs can be excited at this frequency in the same medium, unlike in other hybrid structures such as Pt/Co/YIG~\cite{chen21}. Moreover, the SWs generated in PMA materials are forward-volume SWs, which propagate isotropically~\cite{sushruth20} compared to backward-volume and Damon-Eshbach modes in in-plane magnetised systems, while maintaining high group velocities.

An interesting analogy of our skyrmion lattice as SW emitters can be made with arrays of spin-torque nano-oscillators, which can generate propagating SWs and get synchronised via them~\cite{sani13}. These structures have been proposed for implementing unconventional computing schemes~\cite{macia11}. We have shown here that SWs can be emitted in the uniform background by the skyrmion dynamics, but one also expects that in return, SWs can influence the skyrmion dynamics~\cite{schuette14}, which could enlarge the processing capabilities of this type of devices.

Our study also paves the way toward creating reconfigurable magnonic crystals, since the skyrmion lattices we observe are quasi-periodic and dynamically tunable~\cite{garst17}. We can get a glimpse of the role of this lattice in the observation of SW interference patterns in the simulated HFM. However, the properties of the skyrmion lattice as a magnonic crystal also needs to be explored using broadband propagating SW spectroscopy \cite{ciubotaru16} or advanced SW imaging methods \cite{wintz16, merbouche21}. Lastly, our system could serve as a platform to study magnon-skyrmion scattering \cite{schuette14}, emergent magnon motifs \cite{watanabe20}, or chiral magnonic edge states \cite{diaz20}.

% Not sure if these paragraphs are useful.

\section*{Methods}

\subsection*{Experimental details}
Ta(10)/Pt(8)/[FeCoB(1.2)/AlO$_x$(1.0)/Pt(1.6)]$_{19}$/FeCoB(1.2)
/AlO$_x$(1.0)/Pt(3) multilayer stacks (where FeCoB $\equiv$ Fe$_{70}$Co$_{10}$B$_{20}$, and thicknesses given in nm) were deposited by dc magnetron sputtering at room temperature on thermally-oxidised silicon substrates, under Ar gas flow at a pressure of 0.25 Pa. The base pressure of the sputtering equipment was $5 \times 10^{-6}$~Pa. magnetisation cycles were measured by alternating gradient force magnetometry on $3 \times 3$~mm$^2$ films. MFM images were obtained at room temperature and ambient pressure using a double-pass lift mode, detecting the phase shift (MFM signal) of the second pass after a topographic measurement. The employed low moment tips were home made~\cite{legrand18}. Small permanent magnets with tunable gap and calibrated values of field were used to bias the samples while imaging them. VNA-FMR was conducted on $2 \times 4$~mm$^2$ films facing the $50~\mu$m wide central conductor of a 50~$\Omega$ matched coplanar waveguide. The OP magnetic field was ramped up and down between $-550$~mT and $+550$~mT. For each of the 400 field steps, the VNA frequency was swept from  0.1~GHz to 20~GHz with a number of points set to 1491 and a sweep time of {45~s}. To improve the signal to noise ratio, the data were averaged over 30 magnetic field cycles. To remove the field independent background and normalise the transmission signal data $S_{21}$, we calculate $\mathrm{d_D}S_{21}$ using the derivative divide method~\cite{maier-flaig18}. To increase the contrast further, we apply an additional frequency differentiation, and plot the imaginary part of the obtained quantity in Fig.~\ref{fig:FMRmodes}(a).

\subsection*{Extraction of magnetic parameters}
The magnetisation of the studied multilayers was determined by SQUID magnetometry. The 1.2~nm thick FeCoB is found to have a saturation magnetisation $M_s=1.20 \pm 0.1$~MA/m. Using angle dependent cavity-FMR, the gyromagnetic ratio $\gamma_0/(2\pi) = 28.75 \pm 0.10$~GHz/T and the total uniaxial anisotropy field including demagnetising and PMA contributions $\mu_0 (M_s-H_k) = 354.7 \pm 0.5$~mT are determined (see \textcolor{blue}{Supplementary Note 1}). In the case of uniaxial anisotropy $\mu_0 H_k = 2 K_u/M_s$, hence $K_u=0.69 \pm 0.11$~MJ/m$^3$. By assuming $M_s=1.2$~MA/m and $K_u=0.7$~MJ/m$^3$, the DMI and exchange parameters were iteratively determined by adjusting the period of the simulated labyrinthine domains to the one ($\simeq 170$~nm) observed by MFM at remanence \cite{legrand18}. We find the DMI constant to be $D=1.2$~mJ/m$^2$ for an exchange constant $A=15$~pJ/m. 
%These values were used in subsequent micromagnetic simulations which satisfactorily reproduce the magnetisation curves, MFM images, and FMR data.

\subsection*{Determination of skyrmion diameter}
From the MFM measurements of the quasi static magnetic configuration as a function of the applied field, we extract skyrmion density by using a MATLAB program which uses binarisation. However, as the image resolution is not good enough to estimate the skyrmion diameters directly from the MFM images (which would anyway require to know the transfer function of the tip \cite{feng22}, and might be complicated by the influence of the tip's stray field), we extract the apparent skyrmion diameters ($2R_\mathrm{sk}$) indirectly, for the field values where the skyrmion density ($d_\mathrm{sk}$) is close to maximum, using the out-of-plane experimental magnetisation curve: $R_\mathrm{sk} = \sqrt{(1-m_z)/(2\pi d_\mathrm{sk})}$. In Fig.~\ref{fig:loopMFM}(c) the error bars constitute the errors on the skyrmion density (proportional to the density and inversely proportional to the image size: $\sim 8$\%) and magnetic field values to account for the slight misalignment of the magnetic pole and the magnetic tip in the MFM setup (10\%). In the case of simulated MFM images, we used the same MATLAB program that binarises the simulated MFM images, detects and finds the diameters of all the skyrmions, and calculates the mean and the standard deviation. For the simulations, we also compared the skyrmion diameter i/ calculated by the program and ii/ estimated from skyrmion density and $m_z$ curve as a function of field when the skyrmion density is close to maximum. The small difference in the skyrmion diameters for the two methods lies within the standard deviation of the former method, indicated by the shaded area in grey in Fig.~\ref{fig:Simstat}(b).

\subsection*{Micromagnetic simulations}
Simulations were carried out with the open-source finite-difference micromagnetics code \textsc{Mumax3}~\cite{vansteenkiste14}. 
We used values of $M_s=1.3375$~MA/m and $K_u=0.9$~MJ/m$^3$, which are slightly larger than the experimentally-determined values but were chosen to reproduce quantitatively the mode frequencies (see \textcolor{blue}{Supplementary Figure 12}). The geometry consists of a rectangular slab of volume $1~\mu\mathrm{m} \times 1~\mu\mathrm{m} \times 72$~nm discretised using $128 \times 128 \times 60$ rectangular cells for the susceptibility calculations and a volume of $2~\mu\mathrm{m} \times 2~\mu\mathrm{m} \times 72$~nm with $256 \times 256 \times 60$ cells for the quasi-static hysteresis loop calculations. The simulated stack consists of 20 ferromagnetic layers with thickness 1.2~nm separated by 2.4~nm empty spacers, which accurately mimics the experimental stack, in which the total thickness of non-magnetic Pt and AlO$_x$ layers separating adjacent FeCoB layers is 2.6~nm. Periodic boundary conditions are used in $x$ and $y$ directions in the film plane. The code performs a numerical time integration of the Landau-Lifshitz-Gilbert equation for the magnetisation dynamics,
\begin{equation}
\frac{d\mathbf{m}}{dt} = -|\gamma_0| \mathbf{m} \times \mathbf{H}_\mathrm{eff} + \alpha \mathbf{m} \times \frac{d\mathbf{m}}{dt},
\label{eq:LLG}
\end{equation}
where $\|\mathbf{m}(\mathbf{r},t) \| = 1$ is a unit vector representing the orientation of the magnetisation field, $\gamma_0 = \mu_0 g \mu_B / \hbar$ is the gyromagnetic constant, and $\alpha$ is the Gilbert damping constant. The effective field, $\mathbf{H}_\mathrm{eff} = -(1/\mu_0M_s)\delta U/\delta \mathbf{m}$, represents a variational derivative of the total magnetic energy $U$ with respect to the magnetisation, where $U$ contains contributions from the Zeeman, nearest-neighbour Heisenberg exchange, interfacial Dzyaloshinskii-Moriya, and dipole-dipole interactions along with a uniaxial anisotropy with an easy axis along $\hat{\mathbf{z}}$, perpendicular to the film plane. Note that while the lateral cell size of 7.8125 nm used for the simulations presented here is larger than the characteristic exchange length, $l_\mathrm{ex} = \sqrt{2A/\mu_0 M_s^2} \simeq 3.65$ nm, we have verified that using smaller discretisation cells do not affect the salient features presented here (see \textcolor{blue}{Supplementary Figure 13}).

To simulate quasi-static processes such as hysteresis loops, we begin with a uniform magnetic state oriented along $-z$ under an applied field of $\mu_0 H_z = -500$~mT. A Langevin-dynamics simulation with $T = 300$ K is then performed over 1 ns to mimic experimental conditions in which thermal fluctuations help to sample different metastable states in the energy landscape. The system is then relaxed without the precessional term in Eq.~(\ref{eq:LLG}) toward an energy minimum. This procedure of running Langevin dynamics, followed by relaxation, is then repeated two more times. After this, the micromagnetic configuration is recorded and assigned as the equilibrium state for the applied field value. This equilibrium state then serves as the initial state for the next value of the applied field, which is increased to $\mu_0 H_z = 500$~mT in increments of 1~mT. The simulated MFM images are computed from the equilibrium state by assuming a sample-to-tip distance of 50 nm.

To simulate the microwave susceptibility of the FeCoB multilayer, we compute the response of the magnetisation to harmonic excitation fields. While it is common practice (and usually more efficient computationally) to calculate the eigenmode spectra from the Fourier transform of the transient response to pulsed fields, such an approach for the present system results in strong artifacts arising from non-resonant drift of the background magnetic state, such as the displacement of the skyrmion lattice or labyrinthine domain structure. Instead, we adopt an approach that more closely resembles the experimental protocol. For each value of the external field $H_z$, we compute the response of the precomputed static configuration, $m_0(\mathbf{r})$, to sinusoidal in-plane fields with frequency $f$, $h_x=h_\mathrm{rf} \sin\left(2\pi f t\right)$, over 20 periods $\tau=1/f$, where the spatial average of the resulting component $m_x(t)$ is recorded at intervals of 0.1 $\tau$. In this procedure, the Gilbert damping parameter is set to $\alpha=0.02$ and the excitation amplitude to $\mu_0 h_\mathrm{rf}=0.1$~mT (1 mT in Fig.~\ref{fig:SimHFM} and \textcolor{blue}{Supplementary Figure 12}). We estimate the imaginary part of the susceptibility $\chi$ at this frequency, $\mathrm{Im}[\chi(f)]$, by projecting $\langle m_x(t) \rangle$ onto $-\cos\left(2\pi f t\right)$,
\begin{equation}
\mathrm{Im}\left[\chi(f)\right] \simeq - \frac{1}{20 \tau}\int_0^{20\tau} \langle m_x(t) \rangle \cos\left(2\pi f t\right) dt,
\end{equation}
which is the quantity shown in Figs.~\ref{fig:Simdyn}(a) and \ref{fig:Simdyn}(b). $\mathrm{Im}[\chi(f)]$ exhibits peaks at resonance when $m_x$ is in perfect quadrature with respect to $h_x$. This procedure is repeated as a function of $f$ ranging from 0.1~GHz to 20~GHz in steps of 0.1~GHz for the map in Fig.~\ref{fig:Simdyn}(a) and in steps of 0.02~GHz for the data in Figs.~\ref{fig:Simdyn}(b,c). 

The spatial mode profile at a given mode resonance $f_n$ is extracted by driving the system harmonically at $f_n$ with $h_x=h_\mathrm{rf} \sin\left(2\pi f_n t\right)$ and by recording, cell-by-cell, the components of magnetisation at intervals of 0.1 $\tau_n$, as for the calculation of the susceptibility. The mode profiles are characterised by the fluctuation amplitude $\langle \delta m_x(\mathbf{r}) \rangle$, which is computed in an analogous way to the susceptibility, where we drive the system over 100 periods and then compute, after transients are washed out, 
\begin{equation}
\langle \delta m_x(\mathbf{r}) \rangle = -\frac{1}{2\tau_n}\int_0^{2\tau_n} [m_x(\mathbf{r},t)-m_{x,0}(\mathbf{r})] \cos{(2\pi f_n t)} \; dt,
\end{equation}
over two additional periods. Visually, dark blue regions contribute most to the overall $\mathrm{Im}[\chi(f)]$, while equally dark red and blue regions cancel each other out because they represent fluctuations with opposite sign.

\section*{Data availability}
Data are available from the corresponding authors upon reasonable request.

\section*{Acknowledgements}
This work was partially supported by the French National Research Agency (ANR) under grant no. ANR-17-CE24-0025 (TOPSKY) and the Horizon2020 Research Framework Programme of the European Commission under grant nos. 824123 (SKYTOP) and 899646 (k-NET). It is also supported by a public grant overseen by the ANR as part of the ``Investissements d'Avenir'' program (Labex NanoSaclay, reference: ANR-10-LABX-0035).

\section*{Author contributions}
N.R., V.C., T.D., J.-V.K and G.d.L. conceived the project. The samples were designed, grown and characterised by T.S., Y.S., F.A., N.R. and V.C. T.S. performed the MFM experiments with the help of A.V. and K.B. H.H. conducted the cavity-FMR measurements. T.S. and I.N. performed initial FMR characterizations. T.D. conducted the VNA-FMR experiments and analyzed the results with T.S. and G.d.L. J.-V.K. performed the micromagnetic simulations and analyzed them with T.S. and G.d.L. T.S., J.-V.K. and G.d.L. wrote the manuscript. All the authors discussed the data and commented on the manuscript.

\section*{Competing interests}
The authors declare no competing interests.

\section*{Additional information}

\noindent \textbf{Supplementary information} The online version contains supplementary material available at https://doi.org/xxxxx.

\noindent \textbf{Correspondence} and requests for materials should be addressed to T.S., J.-V.K. or G.d.L.

\bibliography{skyrmFMR}

\end{document}

\title{Supplementary Information --- \\ ``Resonant dynamics of skyrmion lattices in thin film multilayers: Localised modes and spin wave emission''}

% [JVK] Prefer macros defined by RevTeX to define authors and affiliations
\author{Titiksha Srivastava}
\email{titiksha.srivastava@cea.fr}
\affiliation{\UMPhy}
\affiliation{\SPEC}
\author{Yanis Sassi}
\affiliation{\UMPhy}
\author{Fernando Ajejas}
\affiliation{\UMPhy}
\author{Aymeric Vecchiola}
\affiliation{\UMPhy}
\author{Igor Ngouagnia}
\affiliation{\SPEC}
\author{Herv\'e Hurdequint}
\affiliation{\SPEC}
\author{Karim Bouzehouane}
\affiliation{\UMPhy}
\author{Nicolas Reyren}
\affiliation{\UMPhy}
\author{Vincent Cros}
\affiliation{\UMPhy}
\author{Thibaut Devolder}
\affiliation{\CNN}
\author{Joo-Von Kim}
\email{joo-von.kim@c2n.upsaclay.fr}
\affiliation{\CNN}
\author{Gr\'egoire de Loubens}
\email{gregoire.deloubens@cea.fr}
\affiliation{\SPEC}

\maketitle

\tableofcontents

\clearpage
\newpage

%%%%%%%%%%%%%%%%%%%%%%%%%%%%%%%%%%%%%%%%%%%%%%%%%%%%%%%%%%%%%%%%%%%%%%%%%%%%%%%
\section*{Supplementary Note 1. Additional FMR measurements in the saturated state}

\subsection*{Angle dependent cavity-FMR}

Cavity-FMR measurements are performed at a fixed frequency of 9.725~GHz to probe the FMR of the multilayer sample as a function of the angle $\theta_H$ between the applied field and the film plane. Out-of-plane (OP) and in-plane (IP) configurations correspond to $\theta_H=0^\circ$ and $\theta_H=90^\circ$, respectively. The resonant field measured at $\theta_H=0^\circ$ is higher than at $\theta_H=90^\circ$, meaning that the overall anisotropy is in-plane. From the values of the resonant field in these two configurations, one can extract the gyromagnetic ratio $\gamma_0/(2\pi) = 28.75$~GHz/T and the total uniaxial anisotropy field including demagnetising and PMA contributions $\mu_0 (M_s-H_k) = 354.7$~mT. Using these values, the full angular variation of the resonance field is accurately reproduced by the theoretical expectation \cite{hurdequint02}, as shown by the continuous line in Supplementary Figure~\ref{fig:cavity-FMR}(a). The angular dependence of the full width $\Delta H$ at half maximum of the resonance peak is displayed in Supplementary Figure~\ref{fig:cavity-FMR}(b). The minimal value of 33.2~mT is observed in the IP configuration, while the linewidth is only slightly larger in the OP configuration, $\Delta H= 35.5$~mT; it is rather low for this type of multilayer film, which evidences its good dynamical quality.

\subsection*{Damping parameter}

To separate the intrinsic and extrinsic contributions to the FMR linewidth, broadband-FMR measurements are performed in the IP configuration. At 10~GHz, the extracted linewidth is 32.5~mT, in good agreement with the cavity-FMR results. The frequency dependence of $\Delta H$ is plotted in Supplementary Figure~\ref{fig:cavity-FMR}(c) together with its linear fit. The latter yields the frequency independent contribution to the linewidth, found to be about 18~mT, which indicates a fair amount of inhomogeneities in the film. The frequency linear part of the linewidth yields the Gilbert damping parameter of the film, $\alpha=0.022\pm0.003$.

%%%%%%%%%%%%%%%%%%%%%%%%%%%%%%%%%%%%%%%%%%%%%%%%%%%%%%%%%%%%%%%%%%%%%%%%%%%%%%%

\clearpage
\newpage
\section*{Supplementary Note 2. Analysis of the high frequency mode}

\subsection*{Internal field in skyrmion lattice phase}
The spatial profile of the total internal field (\textit{i.e.}, the sum of the applied, anisotropy, and dipolar fields) can be obtained from micromagnetic simulations. The map obtained at 300~mT is shown for layer 12 in Supplementary Figure~\ref{fig:Bi-dispersion}(a). It shows that sufficiently far away from the skyrmions, the static internal field is rather homogeneous in the background domains, around 110 to 120~mT. Due to the strong dipolar fields in the vicinity of the oppositely magnetised skyrmion cores, the internal field substantially increases while approaching the skyrmion boundaries, as shown by the closely spaced isocontours. Inside the skyrmion cores it becomes negative, about $-200$~mT.

\subsection*{Contributions to the eigenfrequency}
As shown by Fig.~5 of the main text the high frequency mode (HFM) corresponds to a precession of the magnetisation localized inside the skyrmion cores. Its frequency can be calculated as the convolution of its spatial profile with a frequency operator taking into account all the contributions to the magnetic energy (Eq.~8 in ref. \cite{naletov11}): Zeeman, anisotropy, dipolar and exchange energies. Even for homogeneously magnetised non-ellipsoidal bodies, it is difficult to evaluate the magnetic self-interactions (magneto-dipolar and inhomogeneous exchange) contributing to the eigenfrequency. As shown in Fig.~3 of the main text, the static configuration in the 20 dipolarly coupled magnetic layers is quite complex in the skyrmion phase. However, to roughly estimate the HFM frequency, one can consider the averaged value of the static internal field inside the skyrmion core where the precession is localized, to which the dynamic magneto-dipolar self-interaction (\textit{i.e.} the depolarization energy of the mode profile on itself) and the exchange contributions should be added. The averaged value $\mu_0 H_{\mathrm{eff},0} \approx -200$~mT of the internal field in the core translates into a frequency of $5.8$~GHz, which is less than the HFM frequency of 15.5~GHz determined by micromagnetic simulations. The exchange contribution to the frequency is given by $(\gamma_0 A k_\mathrm{HFM}^2)/(\pi M_s)$ where $k_\mathrm{HFM}$ is the effective wavevector of the mode. Assuming an effective wavelength of the order of the skyrmion diameter $\approx 100$~nm, one obtains an exchange contribution of 2.6~GHz. The evaluation of the contribution due to the dynamic magneto-dipolar self-interaction is more difficult, because it requires the knowledge of static and dynamic magnetisation profiles to perform the required integration. However, the missing 7.1~GHz fall in the frequency range of this contribution in submicron samples \cite{naletov11}.

\subsection*{Characteristics of the spin waves emitted by the skyrmion cores}
To find the characteristics of the spin waves (SWs) emitted by the precession of the skyrmion cores in the surrounding background domains, one should consider that the latter are homogeneously magnetised OP along the applied field and use the dispersion relation of forward-volume waves \cite{kalinikos86}:
\renewcommand{\theequation}{S1}
\begin{equation}
  \omega (k) = \sqrt{ \left( \omega_{H_{\mathrm{eff},0}} + \frac{2A\gamma_0}{M_s} k^2 \right) \left( \omega_{H_{\mathrm{eff},0}} + \frac{2A\gamma_0}{M_s} k^2 + \omega_M \left( 1 - \frac{1-e^{-kt}}{kt} \right) \right)} \,,
  \label{eq:MSFVW}
\end{equation}
where $\omega_{H_{\mathrm{eff}}}=\gamma_0 H_{\mathrm{eff},0}$ and $\omega_M = \gamma_0 M$. To take into account the multilayer nature of the film, the dynamic dipolar term is evaluated using an effective film medium of total thickness $t=20\times(t_\mathrm{FeCoB}+t_\mathrm{Pt}+t_\mathrm{AlOx})=76$~nm and magnetisation $M=t_\mathrm{FeCoB}/(t_\mathrm{FeCoB}+t_\mathrm{Pt}+t_\mathrm{AlOx})M_s=0.42$~MA/m. This dispersion relation is plotted in Supplementary Figure~\ref{fig:Bi-dispersion}(b) using $\mu_0 H_{\mathrm{eff},0}=110$~mT, which corresponds to the mean value of the static internal field in the interskyrmion region at the bias field of 300~mT shown in Supplementary Figure~\ref{fig:Bi-dispersion}(a). The skyrmion core precession frequency at this field is equal to 15.5~GHz, as indicated by the horizontal dashed line that crosses the dispersion curve at $k=105~\mathrm{\mu m}^{-1}$, corresponding to a wavelength of $\lambda=59.8$~nm. This value is in excellent agreement with the one directly determined from the SW radiation pattern of Fig.~5(c) in the main text. Moreover, it is also possible to compute the group velocity $V_g = \partial \omega / \partial k$ of the forward-volume wave in the effective film, which is plotted in Supplementary Figure~\ref{fig:Bi-dispersion}(c). The group velocity of the SWs propagating at 15.5~GHz is found to be $V_g=960$~m/s. Based on Eq.~(\ref{eq:MSFVW}), forward volume SWs exhibit almost circular precession in our material, therefore one can use $\tau = 1/(\alpha \omega)$ to evaluate their lifetime. At $\omega/(2\pi)=15.5$~GHz, $\tau=1/(\alpha \omega)=0.513$~ns, which allows to estimate the attenuation length of this forward-volume SW, $L_\mathrm{att}=V_g\tau=493$~nm.

%%%%%%%%%%%%%%%%%%%%%%%%%%%%%%%%%%%%%%%%%%%%%%%%%%%%%%%%%%%%%%%%%%%%%%%%%%%%%%%
%\clearpage
%\newpage
%\section*{Supplementary Video}
%
%
%The Supplementary Video visualizes the four excitation modes shown in Fig.~4 of the main text, which correspond to the dynamic response under an applied field of $B_z = $285~mT to a sinusoidal in-plane excitation field of frequency  1.1~GHz, 1.4~GHz and 2.1~GHz (LFMs), 4~GHz and 4.65~GHz (IFMs), and 15.05~GHz (HFM). For each mode, grayscale images of the three magnetisation components $m_x$, $m_y$, and $m_z$ in layer 10 are recorded at a rate of 16 frames per period of the excitation field frequency \textbf{[Check these numbers with the new movies]}. This stroboscopic approach produces animations of equal duration and allows us to compare the dynamics across the different excitation frequencies.

%%%%%%%%%%%%%%%%%%%%%%%%%%%%%%%%%%%%%%%%%%%%%%%%%%%%%%%%%%%%%%%%%%%%%%%%%%%%%%% 

\newpage
\section*{Supplementary Figures}

\begin{figure}[ht]
	\centering
        \includegraphics[width=16cm]{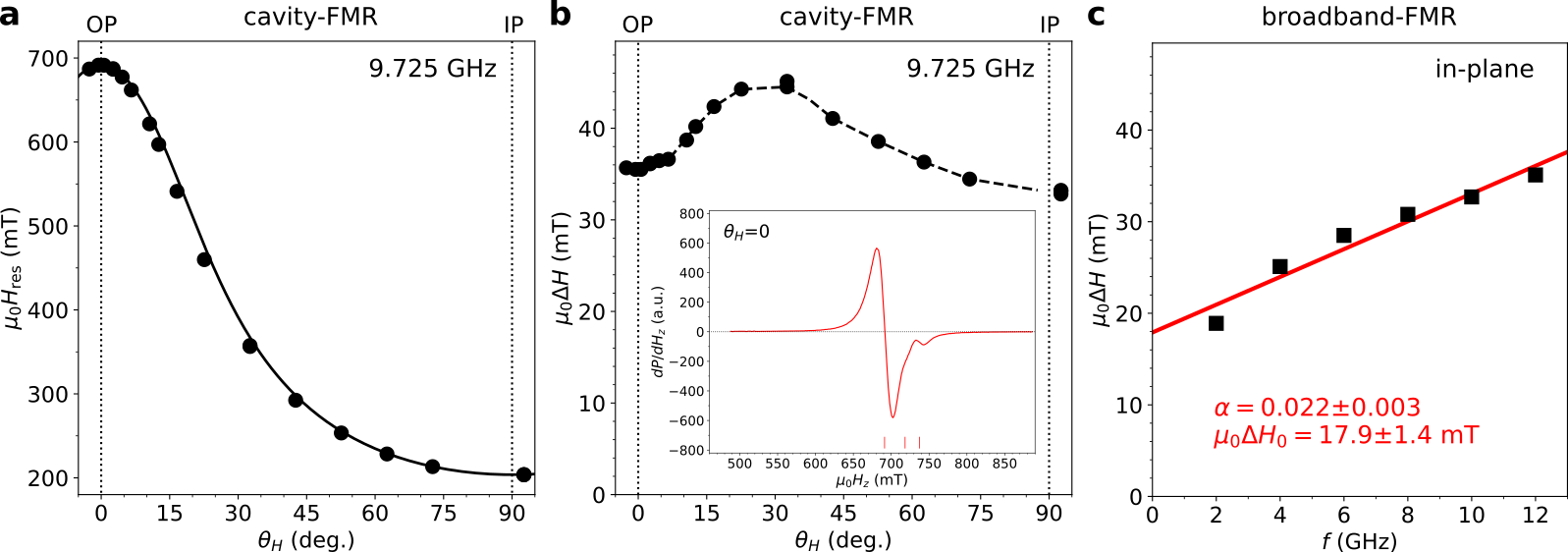}
	\caption[Ferromagnetic resonance in the saturated state]{\textbf{Ferromagnetic resonance in the saturated state}. (a) Evolution of the resonance field as a function of the angle $\theta_H$ between the applied field and the film plane, determined by cavity-FMR at 9.725~GHz. The continuous line is the theoretical expectation using the values of gyromagnetic ratio and total uniaxial anisotropy field extracted from out-of-plane (OP) and in-plane (IP) configurations given in the text. (b) Corresponding dependence of the FMR linewidth on $\theta_H$. The dashed line is a guide to the eye. Inset: spectrum measured in the OP configuration. The resonance field of the uniform Kittel mode is 691.2~mT, additional modes are observed at 718.3~mT and 737.2~mT. (c) Dependence of the in-plane FMR linewidth on frequency measured by broadband FMR. The red line is the linear fit yielding the damping parameter and the inhomogeneous contribution.}
	\label{fig:cavity-FMR}
\end{figure}

\newpage
\begin{figure}[ht]
	\centering
        \includegraphics[width=14cm]{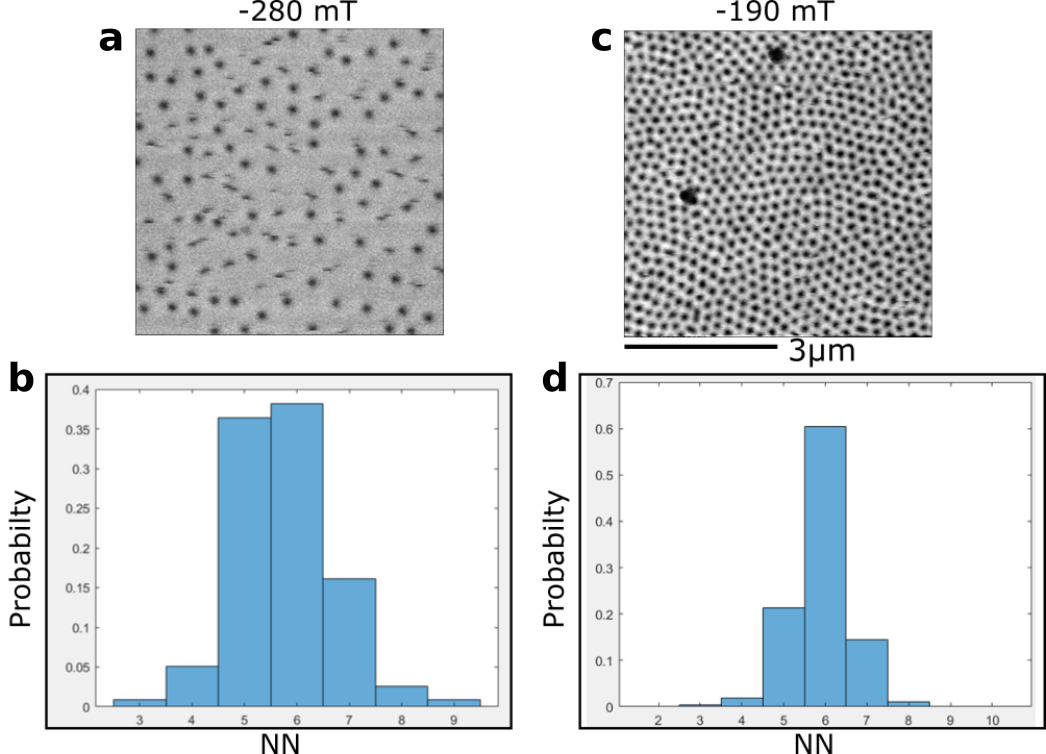}
	\caption[Skyrmion lattice]{\textbf{Skyrmion lattice.} (a) MFM image of skyrmions in disordered state at -280~mT, and (b) nearest neighbour (NN) probability distribution. (c) Skyrmion lattice at -190~mT, and (d) its nearest neighbour probability distribution, demonstrating its almost quasi periodic hexagonal structure.}
	\label{fig:sk-lattice}
\end{figure}

\newpage
\begin{figure}[ht]
	\centering
        \includegraphics[width=16cm]{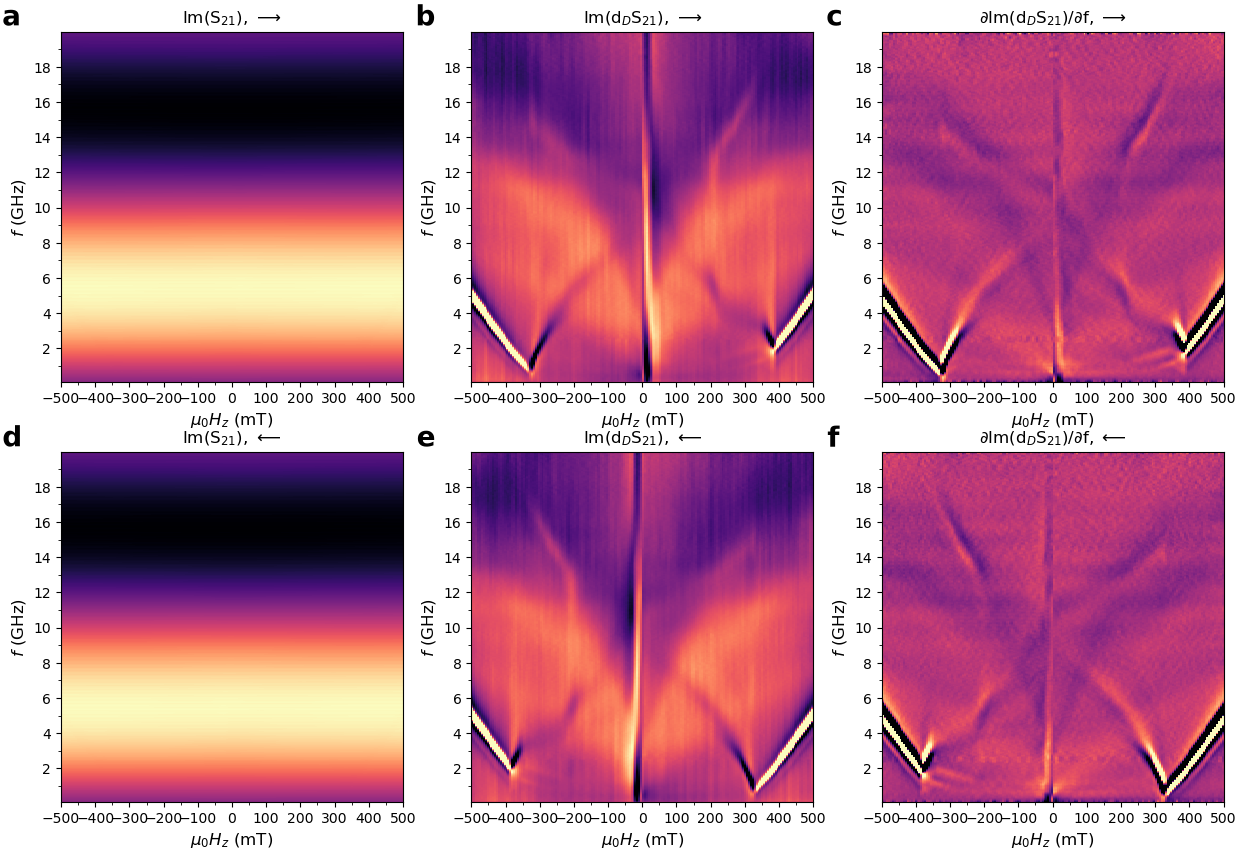}
	\caption[VNA-FMR data processing]{\textbf{VNA-FMR data processing}. (a) Intensity map of the 30-cycle averaged transmission signal $S_{21}$ in the field--frequency coordinates, the field being swept upwards. (b) Map of the $S_{21}$ data processed using the derivative method. (c) Map of the $S_{21}$ data processed using the derivative method followed by a frequency differentiation. (d--f) same as (a--c) for the magnetic field being swept downwards. In addition to the narrow signals ascribed to ferromagnetic resonances in the multilayer sample, there is a very broad background signal coming from the measurement cell which is symmetric between negative and positive field values, as can be seen from the distorted "V" shape in panels (b) and (e), and independent of the sweep direction -- in strong contrast with the resonances originating from the multilayer sample (see also Suppl. Fig.~\ref{fig:hyst-VNA-FMR}). The frequency differentiation attenuates the broad background signal and improves the signal-to-noise ratio of the narrow resonances from the sample, as seen from panels (c) and (f).}
	\label{fig:data-processing}
\end{figure}

\newpage
\begin{figure}[ht]
	\centering
        \includegraphics[width=8cm]{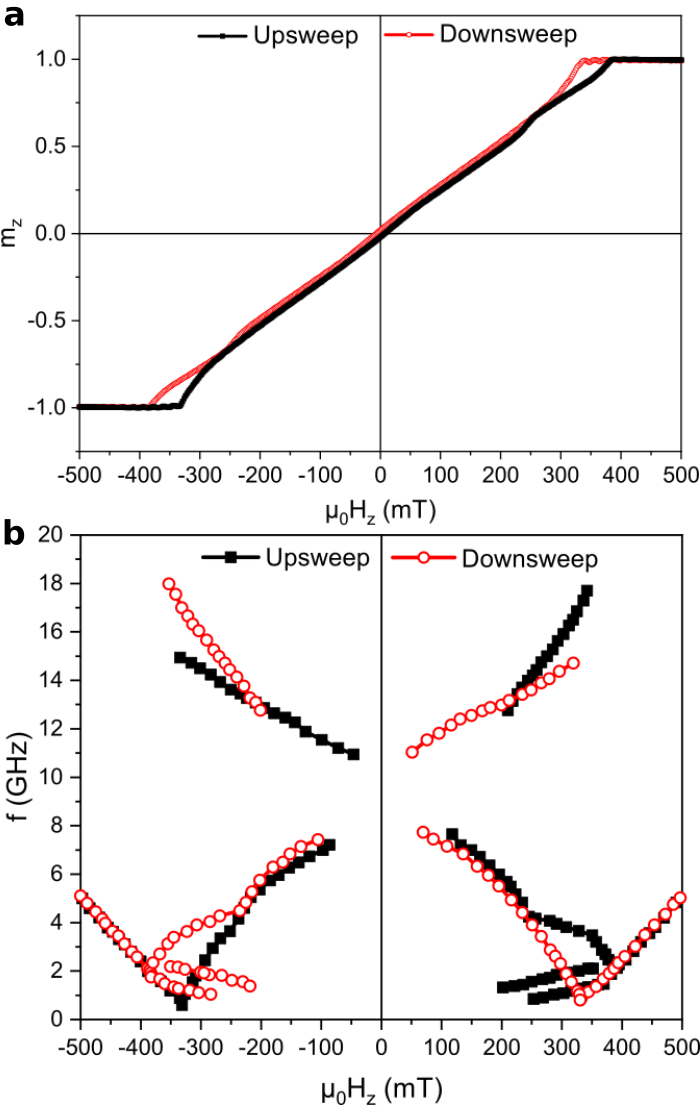}
	\caption[Hysteretic character of the FMR modes]{\textbf{Hysteretic character of the FMR modes}. (a) OP magnetisation curves determined using alternating gradient force magnetometry by sweeping the field up and down. (b) Corresponding dependence of FMR modes determined by broadband VNA-FMR.}
	\label{fig:hyst-VNA-FMR}
\end{figure}

\newpage
\begin{figure}[ht]
	\centering
        \includegraphics[width=13cm]{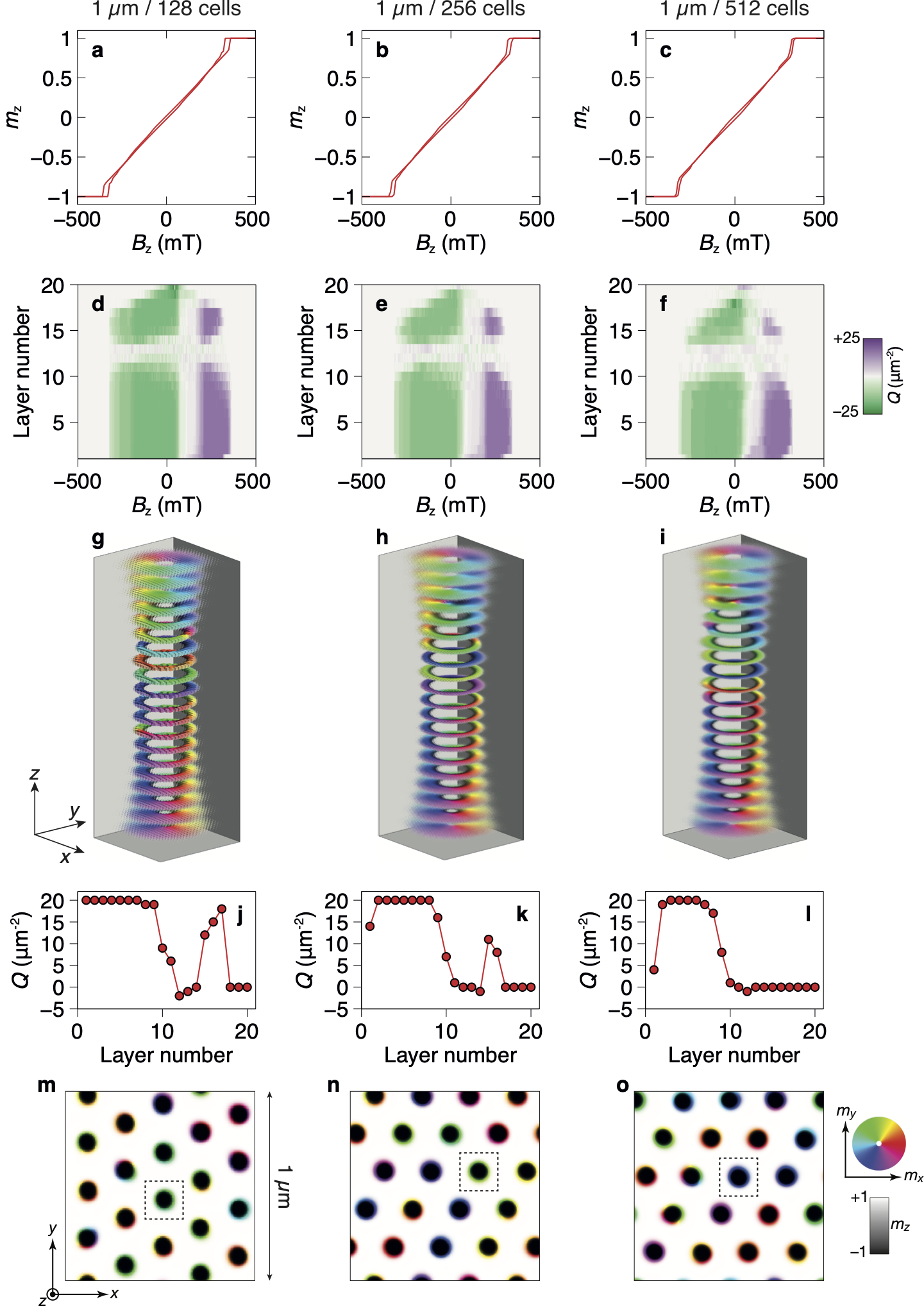}
	\caption[Mesh size dependence of static states]{\textbf{Mesh size dependence of static states}. The simulated static properties, as discussed in Fig.~3 of the main text, are shown for three different discretisation schemes for 1 $\mathrm{\mu m}$ of the simulated geometry: 128 (as discussed in the main text), 256, and 512 cells. (a--c), Hysteresis loops. (d--f), Density plots of the topological charge density per layer as a function of perpendicular applied field $B_z$. (g--i), In-plane components of the magnetisation configuration for a skyrmion at $B_z = 285$ mT. (j--l), Topological charge density as a function of layer number at $B_z = 285$ mT. (m--o), Magnetisation configuration of layer 12 at $B_z = 285$ mT, where the dashed boxes indicate the state shown in (g--i), respectively.}
	\label{fig:meshsize}
\end{figure}

\newpage
\begin{figure}[ht]
	\centering
        \includegraphics[width=9cm]{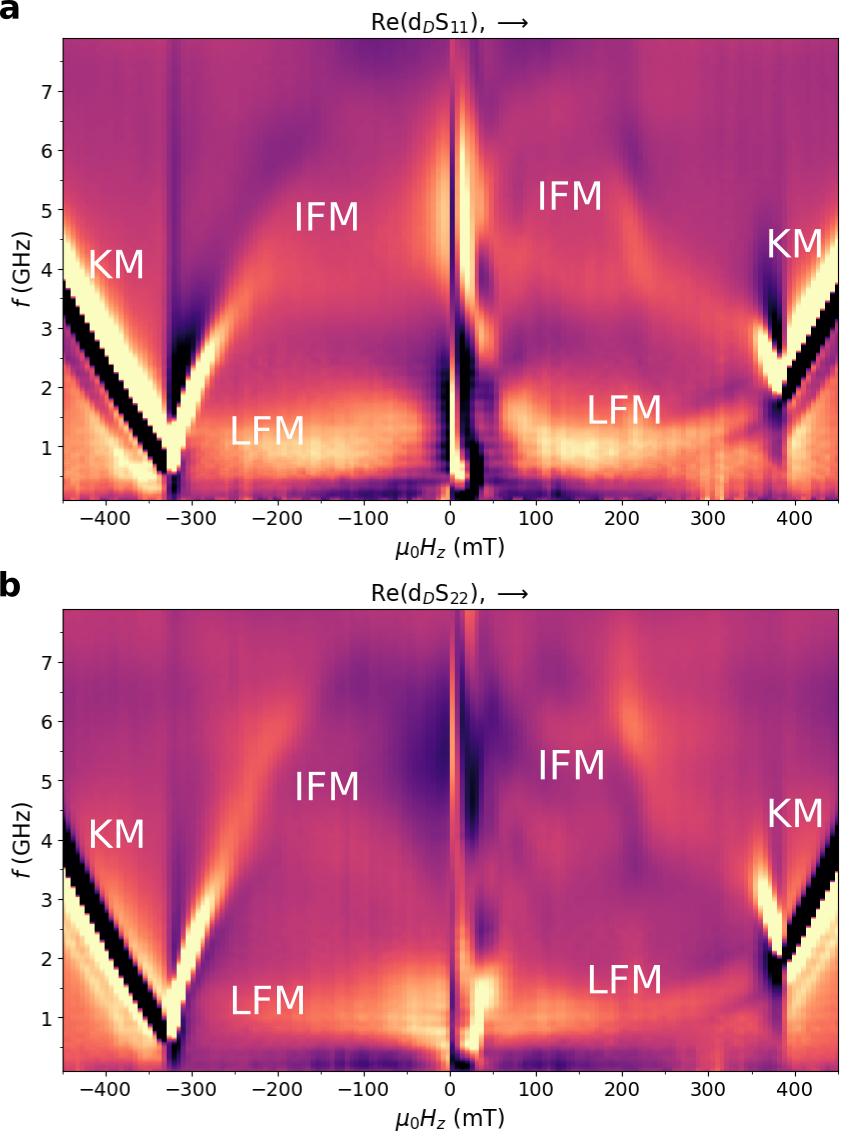}
	\caption[VNA-FMR reflection parameters]{\textbf{VNA-FMR reflection parameters}. Real parts of the (a) $S_{11}$ and (b) $S_{22}$ parameters processed using the derivative divide method. A broad signal between 0.5~GHz and 1.5~GHz is observed on the full field range between $-330$~mT and $390$~mT, which corresponds to the simulated low frequency modes (LFM).}
	\label{fig:S11-S22}
\end{figure}

%\newpage
%\begin{figure}[ht]
%	\centering
%        \includegraphics[width=14cm]{SuppFig7_rev.png}
%	\caption[Layer resolved mode profiles]{\textbf{Layer resolved mode profiles.} Dynamic magnetisation profiles extracted from micromagnetic simulations at 285~mT for the FMR modes of frequencies 1.1~GHz, 1.4~GHz and 2.1~GHz (low frequency modes), 4~GHz and 4.65~GHz (intermediate frequency mode), and 15.05~GHz (high frequency mode) in layers 4, 8, 12 and 16. The profiles in layer 10 of the same four modes are shown in Fig.~4(c) of the main text. For the four different modes, these profiles have similar characteristics in layers 8, 10 and 12 and exhibit a maximum amplitude for layer 12. They vary more strongly in layers 4 and 16, where in fact the static configuration substantially deviates from the one observed in the middle layers due to dipolar effects. The right column displays the layer averaged spatial FFT of the dynamic magnetisation profiles.}
%	\label{fig:layer-modes}
%\end{figure}

\clearpage
\newpage
\begin{figure}[ht]
	\centering
        \includegraphics[width=14cm]{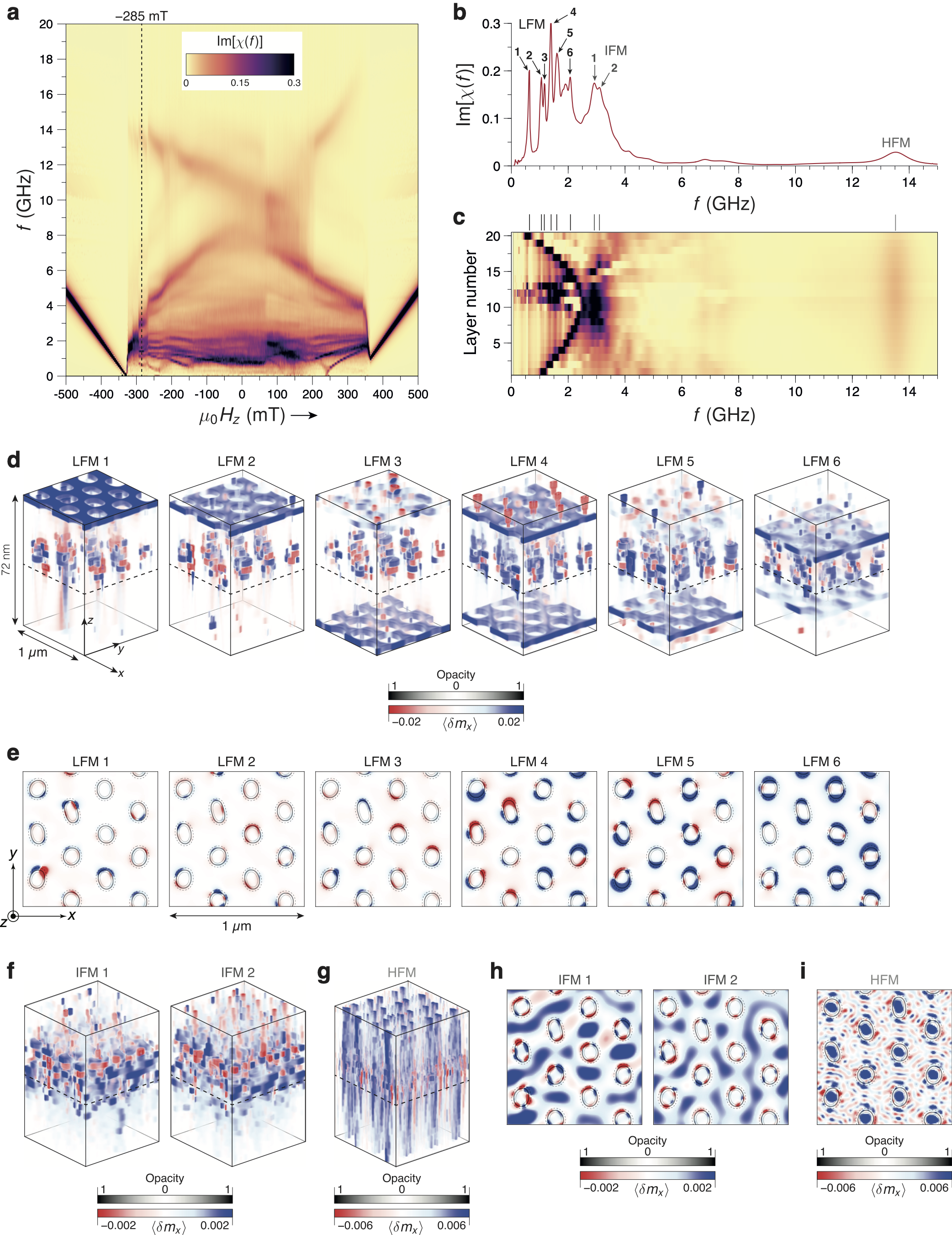}
	\caption{\textbf{Micromagnetic simulations of the dynamic response at $\mu_0 H_z = -285$ mT}. (a) Field-dependent susceptibility map, reproduced from Fig. 4(a). (b) Imaginary part of the magnetic susceptibility at -285~mT, showing six LFM (0.6, 1.06, 1.16, 1.4, 1.6, and 2.08 GHz), two IFM (2.92 and 3.1 GHz) and one HFM (13.52 GHz). (c) Thickness-resolved susceptibility at -285~mT. (d, f, g) Three-dimensional view of the magnetic response $\langle\delta m_x\rangle$ for (d) LFM, (f) IFM, and (g) HFM. (e, h, i) Two-dimensional view of the magnetic response $\langle\delta m_x\rangle$ in layer 10 of the modes in (d,f,g) for (e) LFM, (h) IFM, and (i) IFM, respectively.}
\end{figure}

\clearpage
\newpage
\begin{figure}[ht]
	\centering
        \includegraphics[width=14cm]{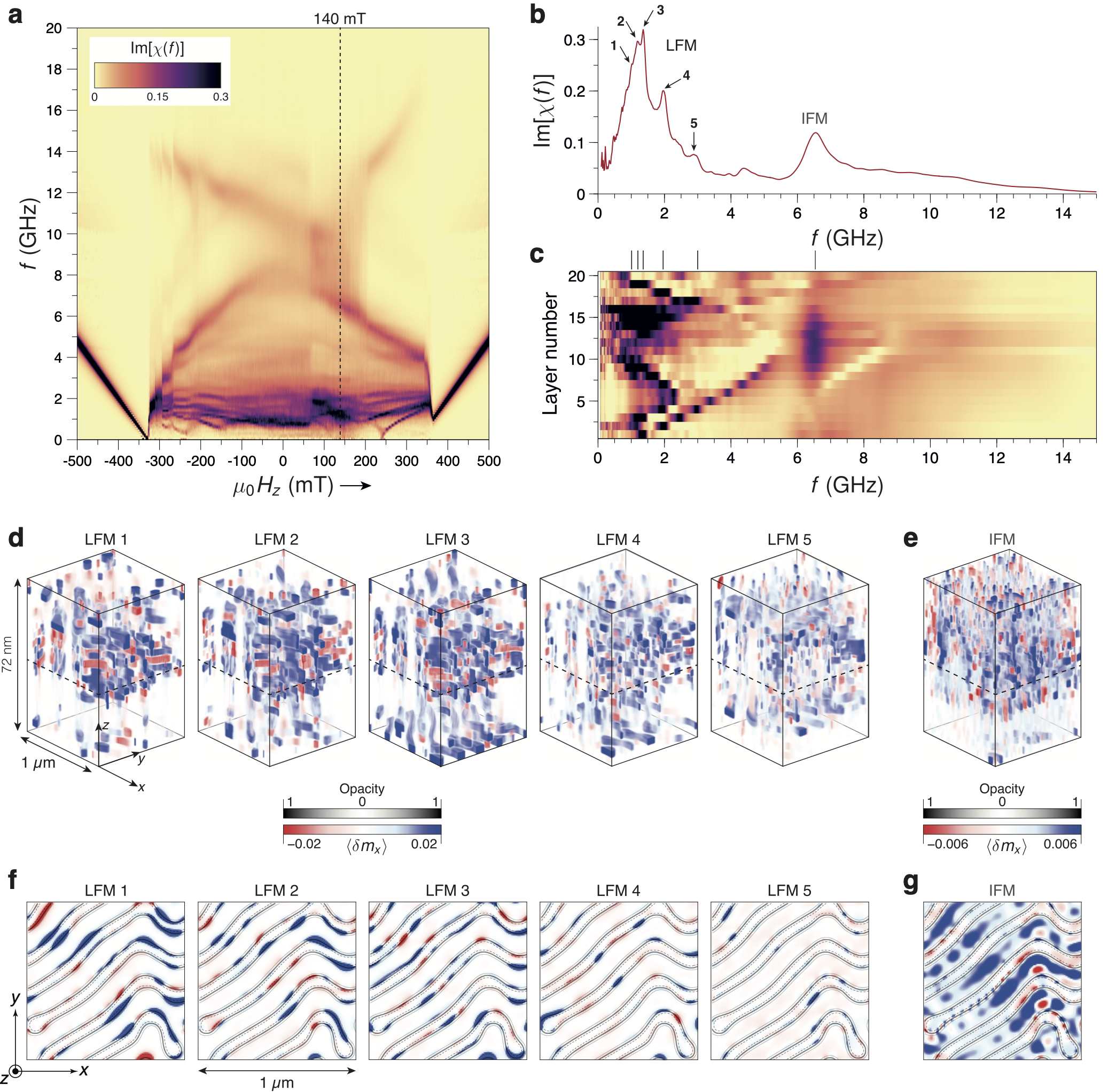}
	\caption{\textbf{Micromagnetic simulations of the dynamic response at $\mu_0 H_z = 140$ mT}. (a) Field-dependent susceptibility map, reproduced from Fig. 4(a). (b) Imaginary part of the magnetic susceptibility at 140~mT, showing five LFM (1.02, 1.2, 1.36, 1.96, and 3 GHz) and one IFM (6.54 GHz). No HFM is observed because no skyrmions are present. (c) Thickness-resolved susceptibility at 140~mT. (d, e) Three-dimensional view of the magnetic response $\langle\delta m_x\rangle$ for (d) LFM and (e) IFM. (f, g) Two-dimensional view of the magnetic response $\langle\delta m_x\rangle$ in layer 10 of the modes in (d,e) for (f) LFM and (g) IFM, respectively.}
\end{figure}

\clearpage
\newpage
\begin{figure}[ht]
	\centering
        \includegraphics[width=16cm]{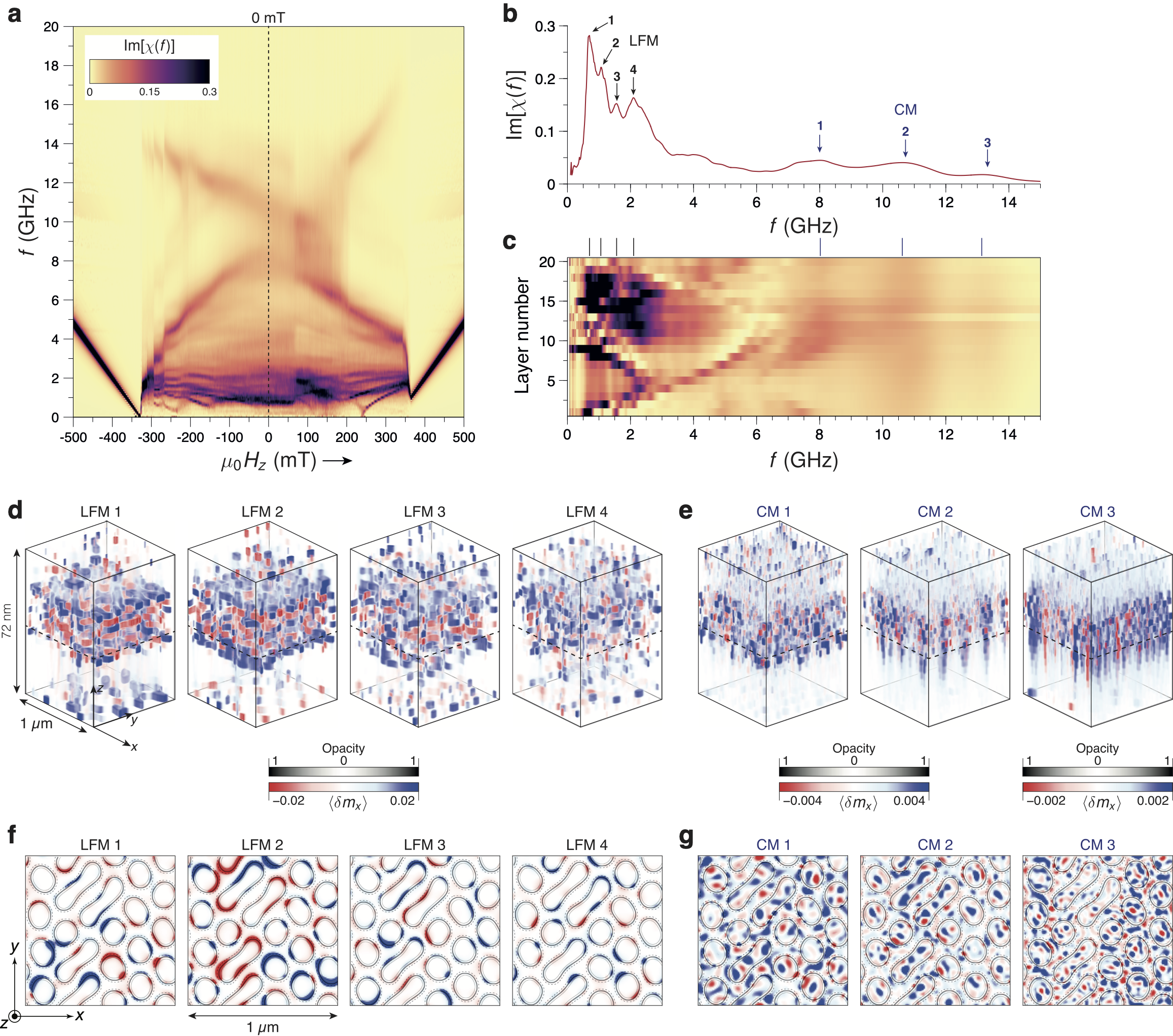}
	\caption{\textbf{Micromagnetic simulations of the dynamic response at $\mu_0 H_z = 0$ mT}. (a) Field-dependent susceptibility map, reproduced from Fig. 4(a). (b) Imaginary part of the magnetic susceptibility at 0~mT, showing four LFM (0.7, 1.06, 1.56, and 2.1 GHz) and three cavity modes (CM) (8.02, 10.62, and 13.14 GHz), where the latter represent spin wave excitations confined within the bubble or elongated stripe domains. (c) Thickness-resolved susceptibility at 0~mT. (d, e) Three-dimensional view of the magnetic response $\langle\delta m_x\rangle$ for (d) LFM and (e) CM. (f, g) Two-dimensional view of the magnetic response $\langle\delta m_x\rangle$ in layer 10 of the modes in (d,e) for (f) LFM and (g) CM, respectively.}
\end{figure}

\clearpage
\newpage
\begin{figure}[ht]
	\centering
        \includegraphics[width=15cm]{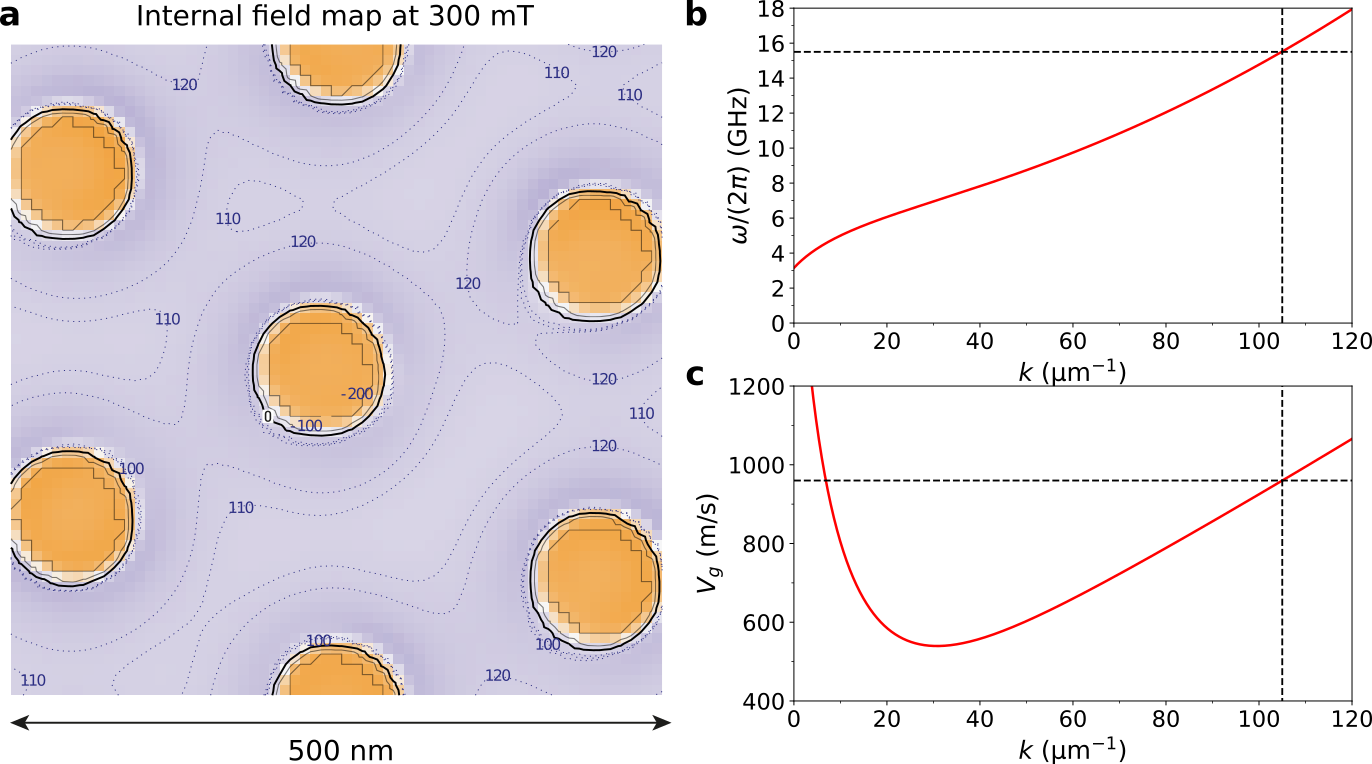}
	\caption[Internal field and spin wave dispersion]{\textbf{Internal field and spin wave dispersion}. (a) Map of the internal field extracted from micromagnetic simulations at 300~mT in layer 12. Values are given in units of mT. (b) Dispersion relation of forward-volume SWs calculated using Eq.~\ref{eq:MSFVW} for a uniform internal field of 110~mT. (c) Corresponding group velocity. The dashed lines in panels (b) and (c) highlight the SWs excited at 15.5~GHz, whose wavelength and group velocity are respectively found to be $\lambda=59.8$~nm and $V_g=960$~m/s.}
	\label{fig:Bi-dispersion}
\end{figure}

\clearpage
\newpage
\begin{figure}[ht]
	\centering
        \includegraphics[width=14cm]{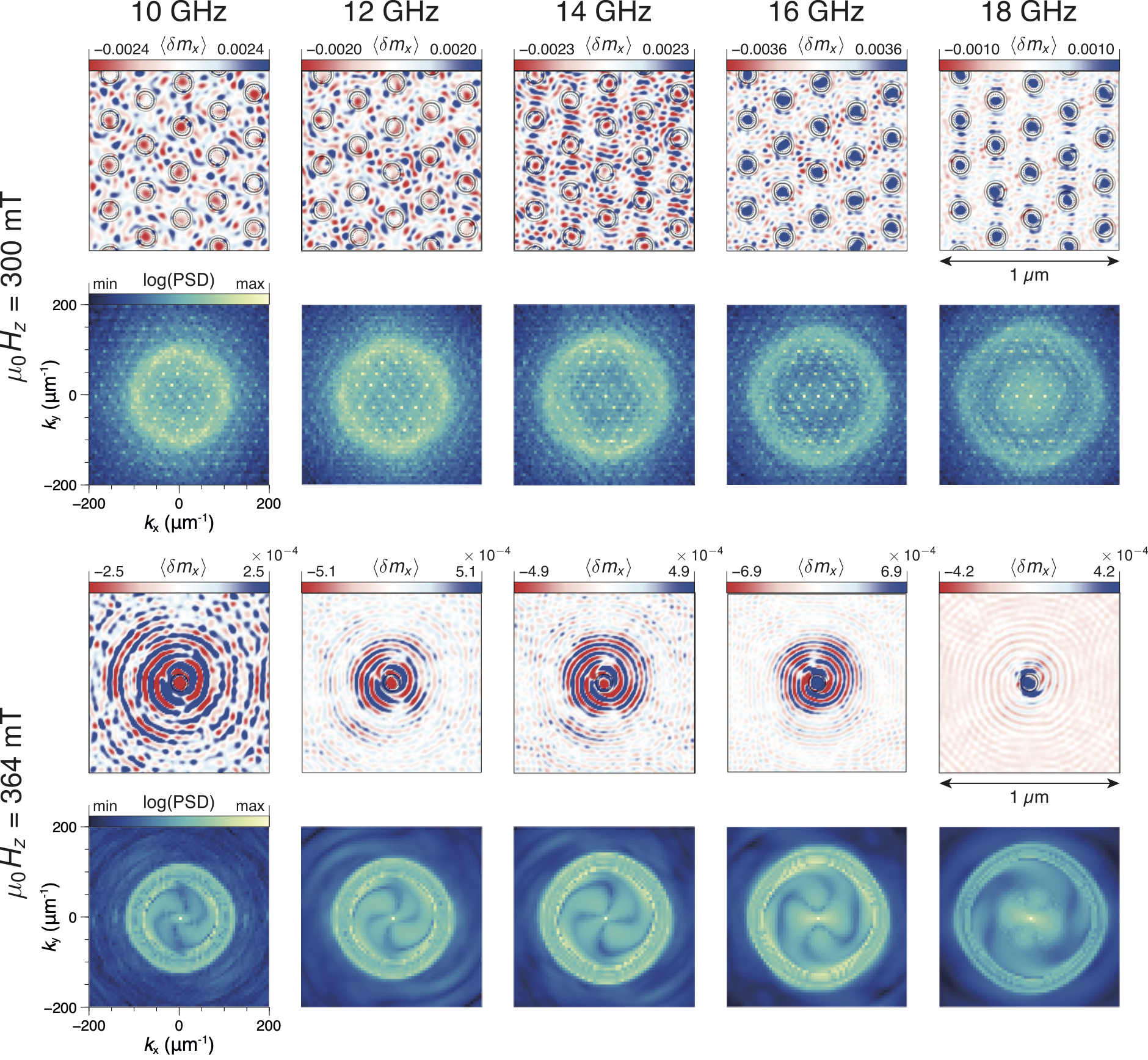}
	\caption[Frequency dependence of the SW interference pattern]{\textbf{Frequency dependence of the SW interference pattern}. Dynamic magnetisation excited in layer 10 at 300~mT (top) and 364~mT (bottom) by an in-plane microwave field of frequency ranging from 10~GHz to 18~GHz, and associated layer averaged spatial FFT. The most efficient frequency to drive the core precession corresponds to the resonance frequency of 15.5~GHz. When using a lower or higher frequency to drive the system, its overall response has a lower amplitude. In addition, following the trend of the dispersion relation shown in Supplementary Figure~\ref{fig:Bi-dispersion}(b), the wave vector of the emitted SWs increases with the excitation frequency of the skyrmion core, as seen from the FFT maps. As a result, the resulting interference patterns also change, in a way that depends on the skyrmion lattice, as shown by the comparison between 300~mT and 364~mT. More generally, the interference patterns depend on the characteristic lengths of the system, which can be tuned by the applied OP field. Both the density and the diameter of the skyrmions indeed depend on the field (see Fig.~3(b) of the main text), which determines the travel length of the SWs between the boundaries of neighbouring skyrmions.}
	\label{fig:SW-freq}
\end{figure}

\clearpage
\newpage
\begin{figure}[ht]
	\centering\includegraphics[width=16cm]{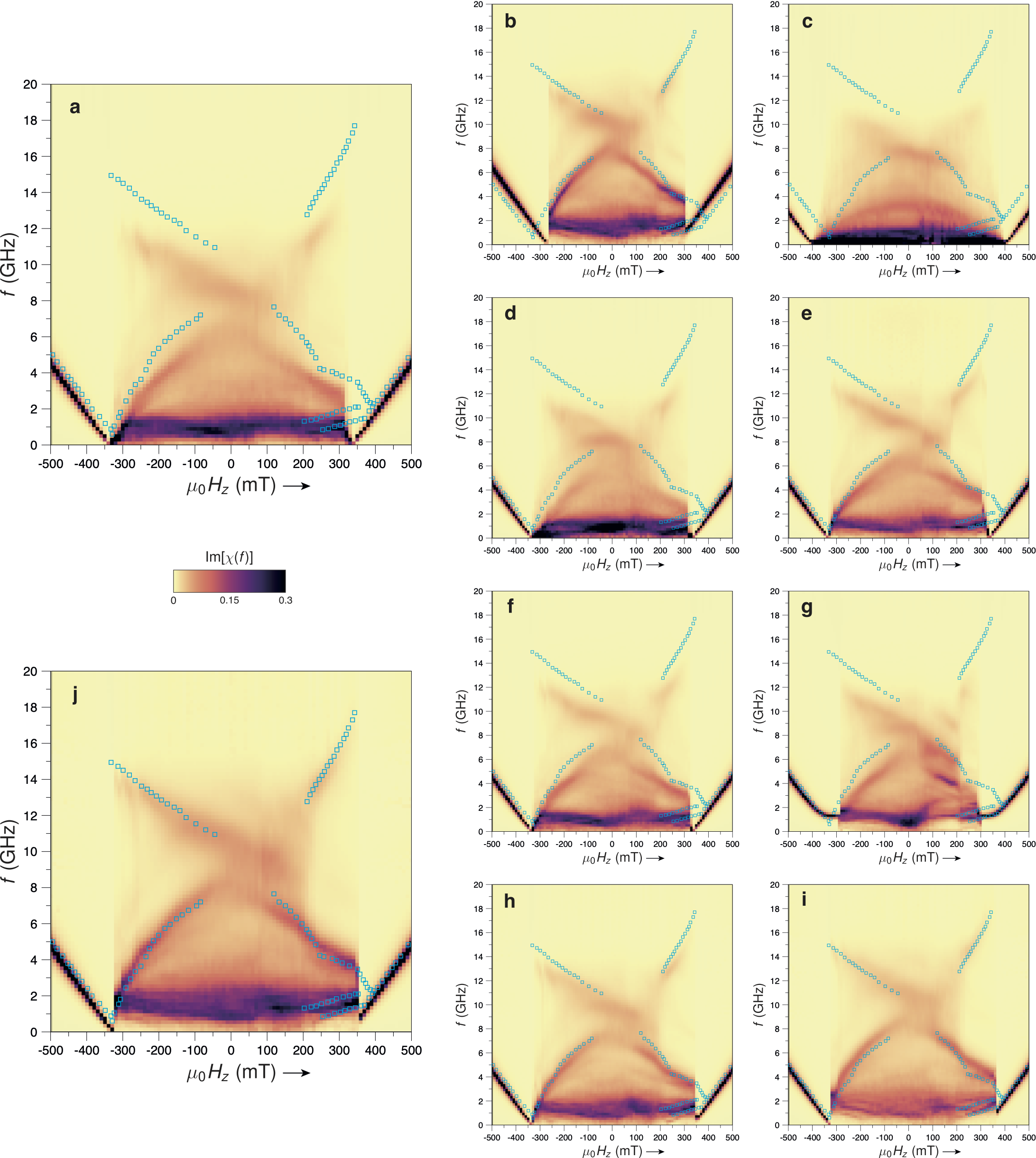}
    \caption[Optimisation of magnetic parameters used in micromagnetic simulations]{\textbf{Optimisation of magnetic parameters used in micromagnetic simulations}. Simulated imaginary part of the dynamic susceptibility, $\textrm{Im}[\chi(f)]$, as a function of applied perpendicular field, $\mu_0 H_z$, for deviations from experimentally-determined values of the micromagnetic parameters, (a) $M_{s,0} = 1.2$ MA/m, $A_0$ = 15 pJ/m, $K_{u,0} = 0.70$ MJ/m$^3$, and $D_0$ = 1.2 mJ/m$^2$. (b) $K_u=K_{u,0} + 0.04$ MJ/m$^3$, (c) $K_u=K_{u,0} - 0.04$ MJ/m$^3$, (d) $A = A_0 + 5$ pJ/m, (e) $A = A_0 - 5$ pJ/m, (f) $D = D_0 + 0.03$ mJ/m$^2$, (g) Experimental values with applied field tilted 2$^\circ$ from $z$-axis along the $x$ direction, (h) $M_s = M_{s,0}+0.1$ MA/m and $K_u=K_{u,0} + 0.14$ MJ/m$^3$, (i) $M_s = M_{s,0}+0.175$ MA/m and $K_u=K_{u,0} + 0.26$ MJ/m$^3$, and (j) $M_s = M_{s,0}+0.1375$ MA/m and $K_u=K_{u,0} + 0.20$ MJ/m$^3$, which represents the best fit. The blue dots represent the experimental data.}
	\label{fig:opti-simu}
\end{figure}

\clearpage
\newpage
\begin{figure}[ht]
	\centering\includegraphics[width=14cm]{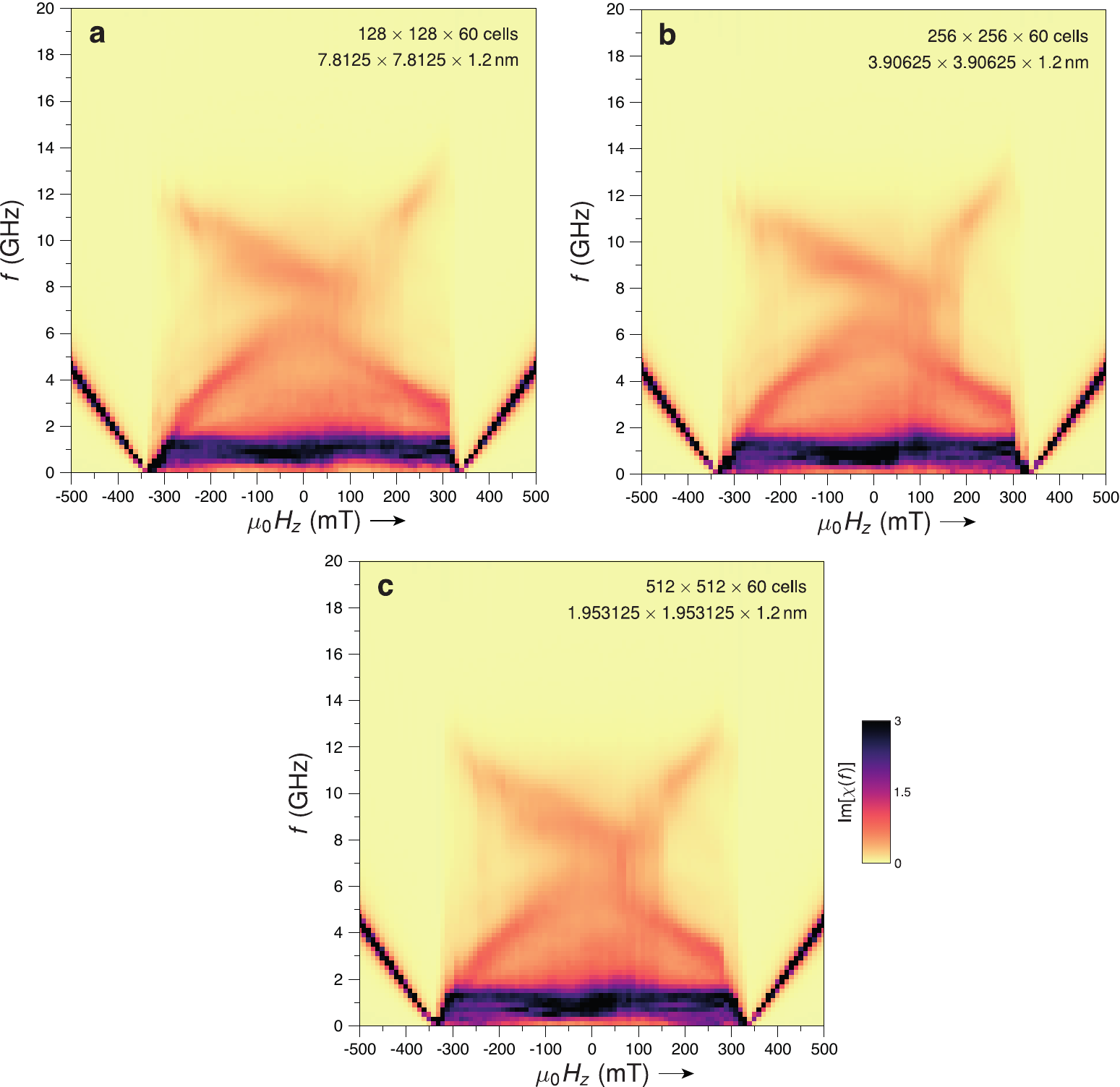}
    \caption[Mesh size dependence of dynamic susceptibility]{\textbf{Mesh size dependence of dynamic susceptibility}. Simulated imaginary part of the dynamic susceptibility, $\textrm{Im}[\chi(f)]$, as a function of applied perpendicular field, $\mu_0 H_z$, for three finite-difference discretization schemes of the $1000 \times 1000 \times 72$ nm simulation box: (a) 128 $\times$ 128 $\times$ 60, (b) 256 $\times$ 256 $\times$ 60, and (c) 512 $\times$ 512 $\times$ 60 finite difference cells. The discretisation scheme and corresponding cell size is indicated on each figure. The micromagnetic parameters used correspond to the experimental values [see Supplementary Figure 12(a).].}
\end{figure}

%%%%%%%%%%%%%%%%%%%%%%%%%%%%%%%%%%%%%%%%%%%%%%%%%%%%%%%%%%%%%%%%%%%%%%%%%%%%%%% 

\clearpage
\newpage
\section*{Supplementary References}

%% Include extra code to remove duplicate TOC entry
\let\oldaddcontentsline\addcontentsline% Store \addcontentsline
\renewcommand{\addcontentsline}[3]{}% Make \addcontentsline a no-op
\bibliography{skyrmFMR}{}
\bibliographystyle{naturemag}
\let\addcontentsline\oldaddcontentsline% Restore \addcontentsline